\documentclass[onefignum,onetabnum]{siamonline190516}
\usepackage{styleFile}
\usepackage{todonotes}
\usepackage{comment}

\ifpdf
\hypersetup{
  pdftitle={Bayesian inference of an uncertain generalized diffusion operator},
  pdfauthor={T. Portone and R.D. Moser}
}
\fi

\title{Bayesian inference of an uncertain generalized diffusion operator\thanks{Submitted to the editors 4/29/21
\funding{
The support of this work by the U.S.~Department of Energy under contracts DE-SC0009286 and DE-SC0019303 is gratefully acknowledged. 
    }
}}
\author{
  T.~Portone\thanks{Optimization and Uncertainty Quantification, Sandia National Laboratories, Albuquerque, NM, 87123 (\email{tporton@sandia.gov}).} \and
R.D.~Moser\thanks{Oden Institute for Computational Engineering and Sciences, University of Texas at Austin (\email{rmoser@oden.utexas.edu}).} 
}

\begin{document}
\maketitle
\begin{abstract}
\pagenumbering{gobble}

This paper defines a novel Bayesian inverse problem to infer an infinite-dimensional uncertain operator appearing in a differential equation, whose action on an observable state variable affects its dynamics. 
  Inference is made tractable by parametrizing the operator using its eigendecomposition.
  The plausibility of operator inference in the sparse data regime is explored in terms of an uncertain, generalized diffusion operator appearing in an evolution equation for a contaminant's transport through a heterogeneous porous medium. 
  Sparse data are augmented with prior information through the imposition of deterministic constraints on the eigendecomposition and the use of qualitative information about the system in the definition of the prior distribution.
Limited observations of the state variable's evolution are used as data for inference, and the dependence on the solution of the inverse problem is studied as a function of the frequency of observations, as well as on whether or not the data is collected as a spatial or time series.
\end{abstract}
\pagenumbering{arabic}



\tableofcontents
\section{Introduction}
In the past decade, a greater understanding of infinite-dimensional Bayesian inference has developed, but it has largely been in the context of inferring functions \cite{hosseini,stuart}. 
On other fronts, Bayesian inference of operators has been considered in several contexts, but the problem is either finite-dimensional or can be recast as a field inversion problem.
For instance, in \cite{levin2009}, the kernel of a convolution operator was inferred along with an image in a blind deconvolution problem, converting operator inference to field inversion.
Much work has focused on inference of the covariance matrix of a multivariate Gaussian distribution, for example to quantify uncertain measurement errors \cite{daniels1999,huang2013}.
These matrices are finite-dimensional and do not directly affect the dynamics of the state, only their presumed measurement error.
Furthermore, in both of these examples, the operator is not affecting the dynamics of the problem.
In \cite{morrison}, an operator affecting the state dynamics was inferred, but the operator was finite-dimensional.

Non-Bayesian methods for inferring an operator affecting state dynamics from observations of the state variable have recently been developed.
For instance, in \cite{peherstorfer2016} the operators in a reduced-order model are inferred deterministically using data generated from a higher-fidelity model's output, taken at a variety of times, locations and model parameter values.
The types of operators in question appear in dynamical systems and are often discretizations of differential operators, but the inference is in a deterministic setting.
In large part, the operator inference problems mentioned here focus on inferring the elements of a finite-dimensional operator's matrix representation directly. 

  In contrast, in this work prior knowledge of invariance in the modeled system (specifically, translation invariance) is exploited to determine the eigendecomposition of an unknown linear operator, and the inference problem is cast in terms of the operator's eigenvalues.
A Bayesian inverse problem is defined to infer the infinite-dimensional differential operator's spectrum using observations of the state variable whose dynamics it affects.
  A favorable property of this approach is that the dimension of the inverse problem does not depend on the discretization of the problem, as is the case when inferring the matrix representation of the discretized operator. 
  Instead the dimension of the inverse problem is determined by the spectral content of the solution to the inverse problem; that is, how many eigenvalues are informed by the observational data as determined by a global sensitivity analysis.
In \cite{pinns,wu2020data} an operator inference is formulated in terms of a differential operator, parametrizing its symbol in Fourier space using a neural network, 
and deterministic constraints based on physical properties are placed on the Fourier symbol.
The problem formulations of \cite{peherstorfer2016,pinns,wu2020data} are most similar to the Bayesian inverse problem posed here, however all methods use the full space- and time-varying evolution of the state variable(s), possibly for multiple initial conditions and/or model parameterizations, to infer the unknown operator.

This work instead focuses on the case of limited data, where either a snapshot in time of the spatially-varying state, or a time-series observation of the state at a specific location is observed. 
Data sparsity is a common issue in realistic physical applications. 
This work is a first step in investigating the feasibility of inferring uncertain or unknown dynamics governing physical phenomena with limited observations. 
  To mitigate the effect of sparse data, physical constraints are imposed deterministically on the operator's formulation.
  Additionally, qualitative information about the behavior of the system is imposed through the prior distributions defined on the operator parameterization.
How the solution to the Bayesian inverse problem depends on the type of data (whether observations are a spatial series or a time series) and on the frequency of observations is explored.
With limited data to constrain the operator, the need to encode prior information into its formulation is especially important. 
How prior information about physical realizability can be encoded in the operator's form is demonstrated here.

The Bayesian inverse problem is focused on inferring an uncertain differential operator representing dispersion in a field-scale model of contaminant transport.
Development of closure models for this phenomenon is an ongoing effort, so this work also has potential applications as a novel method of deriving such closures.
While both practical and theoretical aspects of Bayesian inference of an infinite-dimensional operator must be explored, this work focuses on the practicalities of the problem. 
Conditions under which it is possible to parametrize the inference problem and challenges that arose during the process will be discussed. 

\section{Application problem description}\label{applicationProblemDescription}
When modeling a physical phenomenon, the first step is to bring reliable theory such as conservation laws to bear.
Conservation laws generally contain unclosed terms for which models must be introduced to close the equations.
Often the correct form for such closure models is unknown, given the information available to the modeler. 
This paper recasts the closure problem as a Bayesian inverse problem, where the closure model is represented as an uncertain operator acting on the state variable.

The application problem studied for this work is field-scale transport of a contaminant through a heterogeneous porous medium. 
For the purposes of this discussion, the 2D advection-diffusion equation is considered an accurate representation of the relevant physics, with the velocity governed by Darcy's law and assumed incompressible in a medium with uniform porosity.
For $\mathbf{x}\equiv(x,y)\in[0,L_x]\times[0,L_y]\equiv\Omega$, let
\begin{eqnarray}
    \diffp[]{c(\mathbf{x}, t)}{t} + \div{\ub(\mathbf{x})c(\mathbf{x},t)} = \nu_p \grad^2 c(\mathbf{x},t), \label{eq:detailedConsMass}\\
    \div{\ub} = 0, \label{eq:detailedContinuity}\\
    \ub(\mathbf{x}) = -\kappa(\mathbf{x})\grad p(\mathbf{x}), \label{eq:darcy}
\end{eqnarray}
where $c \in C^{\infty}(\Omega)$ is the concentration field of the contaminant; $\uvec \in C^1(\Omega)$ is the velocity; 
$\nu_p$ represents pore-scale diffusivity; $\kappa(\mathbf{x})$ represents permeability, a measure of how easily fluid travels through the medium; and $p$ is the pressure field.
All fields are assumed periodic in $x$, and $c$ and $\mathbf{u}$ are assumed to satisfy zero-Neumann boundary conditions in $y$.
The initial condition and all parameters are assumed known, except for $\kappa$.
Because the velocity depends on $\kappa$ through Darcy's law, the permeability indirectly determines the transport of the contaminant.
If $\kappa$ were known throughout the entire computational domain, the transport of the contaminant would be completely predictable.

In realistic problems, the structure of a permeability field is not available over the entire  span of the domain due to limitations in sensing technology.
Instead, it is possible to collect samples of the porous medium and study small sections of the domain in a laboratory.
Viewing the permeability as a random field, the samples may be used to determine a mean and correlation structure.
Assuming $\kappa$ is statistically homogeneous---that is, that its statistics do not depend on absolute location---the statistics determined in the lab are representative of its statistics over the whole domain.
Thus, although the detailed behavior of the permeability field is not known, its statistics, and those of the velocity and other quantities that depend on it, can still be predictable.

It is common practice to make such assumptions and perform statistical averaging to derive an equation for the transport of the mean contaminant concentration field \cite{bear2010modeling}, and this approach is taken here. 
Additionally, depthwise ($y$-direction) averaging is performed for two reasons. 
First, although the evolution of the contaminant is assumed to occur in 2D, observations of the contaminant are limited to a depthwise average due to mixing that occurs when drawing fluid from a well for measurement. 
Second, the depthwise variation of the contaminant's concentration is not generally the relevant quantity of interest; of more concern is when the average depthwise concentration of the contaminant exceeds a safe threshold downstream of some contaminant source.

To obtain a set of equations for the statistically- and spatially-averaged concentration, let
\begin{eqnarray*}
  \mean{ f(x,y) } \equiv \frac{1}{L_y}\int_0^{L_y} \mathbb{E}_{\kappa}\left[ f(x,y) \right] \d y
\end{eqnarray*}
for a random field $f$, where $\mathbb{E}_\kappa$ signifies an expectation over the probability space of $\kappa$.
The random field $f$ can thus be written as the sum of its mean and its deviation from that mean:
\begin{eqnarray*}
  f = \mean{f} + f'.
\end{eqnarray*}
Substituting this decomposition of $c$ and $\mathbf{u}=[u,v]$ into the high-fidelity equations \eqref{eq:detailedConsMass} and \eqref{eq:detailedContinuity} and applying the averaging operator to the equations gives
\begin{eqnarray}
  \begin{aligned}
    \diffp[]{\meanc(x,t)}{t} + \mean{u}\diffp[]{\meanc(x,t)}{x} + \diffp[]{\mean{u'c'}(x,t)}{x} = \nu_p \diffp[2]{\mean{c}(x,t)}{x}, \\
     \mean{c}(0,t) = \mean{c}(L_x,t), \\
     \mean{c}(x,0) = c_0(x).
   \end{aligned}
   \label{eq:averaged}
\end{eqnarray}
Note that $\meanc$ is assumed periodic with period $L_x$, 
which enables the computationally-efficient solution of \eqref{eq:averaged} using a Fourier-series expansion.
The periodicity assumption is valid provided the velocity fluctuations are homogeneous with correlation lengths small compared to the period (in this work correlation lengths did not exceed 10\% of $L_x$),
and the contaminant is confined to a region that is small compared to the period (in this work simulations were stopped before the contaminant pulse reached the edges of the domain).
Furthermore, $\mean{u}$ is constant, since by continuity
\begin{eqnarray*}
  \begin{aligned}
    0 = \diffp{\mean{u}}{x} + \diffp{\mean{v}}{y} = \diffp{\mean{u}}{x}.
  \end{aligned}
\end{eqnarray*}
This system of equations is exact but unclosed because of the second-order fluctuating term $\mean{u'c'}$.
This term is often called the dispersive flux, and $\sfrac{\partial \mean{u'c'}}{\partial x}$ is herein called dispersion.
A typical closure model for $\mean{u'c'}$ is gradient diffusion \cite{bear2010modeling}.
However, it is well known that gradient diffusion 
  can be an inadequate model for field-scale transport through heterogeneous media when dispersion dominates pore-scale diffusion and strong heterogeneities induce anomalous diffusion \cite{levy2003, heterogeneousFlow, neuman2009perspective}.
  This work investigates the possibility of deriving a more general closure model for dispersion in scenarios where anomalous diffusion is significant. 
  This is done by defining the problem so that it is in a high-P\'eclet number regime, to ensure dispersion dominates pore-scale diffusion in \eqref{eq:detailedConsMass}.  
  In all cases herein the P\'eclet number is $\text{Pe}\equiv \langle u\rangle  L_y / \nu_p = (1)(1)/(0.01) = 100$.
  Additionally, $\kappa$ are specified with sufficient heterogeneity that anomalous diffusion is observed. 
  Further details on how the problem was defined to produce anomalous diffusion are provided in \cite{portone_thesis}.
  
\subsection{Uncertain operator as closure model}
As an alternative to gradient diffusion, here we pursue the Bayesian inference of an uncertain operator acting on $\mean{c}$ to represent dispersion, defined such that
\begin{eqnarray}
\begin{aligned}
\Lcal \mean{c} &= - \diffp{\mean{u'c'}}{x}.
\end{aligned}
\label{eq:L_def}
\end{eqnarray}
Because $\Lcal$ acts on $\mean{c}$ and appears in its evolution equation, it affects the dynamics of the mean evolution.
The aim of this work is to assess the feasibility of inferring $\Lcal$ using only observations of $\mean{c}$ at varying locations and times.

To enable inference and to encode prior information in the operator's form, $\Lcal$ is parametrized by its eigendecomposition, and relevant physical constraints are enforced on its eigenfunctions and eigenvalues in \Cref{sec:operator_formulation}.
Remaining uncertainties in the operator's parametrization are represented using prior distributions, discussed in \Cref{sec:ip_formulation}.
In \Cref{sec:case_1}, Bayesian inference is performed using data generated from a model for which the operator's form is known \textit{a priori} to assess the feasibility of inferring the operator from data.
Based on the findings of \Cref{sec:case_1}, in \Cref{sec:case_2} the operator is inferred using sample statistics of $\mean{c}$ computed from solutions of the detailed model \eqref{eq:detailedConsMass}-\eqref{eq:darcy}.
The validity of the inferred operator as a closure model for $\sfrac{\partial \mean{u'c'}}{\partial x}$ is also assessed.
Finally, conclusions and future work are discussed in \Cref{sec:conclusions}.

\section{Uncertain operator formulation}\label{sec:operator_formulation}
First, deterministic constraints on the operator's form are imposed to encode prior information and to enable inference via the operator's parametrization.
These constraints are based on physical and mathematical characteristics of the problem that should not be violated by the introduction of the uncertain operator.
For instance, the mean advection-diffusion equation is linear in $\mean{c}$.
It is also shift-invariant because of the statistical homogeneity of the underlying medium.
Finally, it is an expression of conservation of mass. 
The deterministic formulation of $\Lcal$ must be defined to respect these constraints.

To respect linearity in $\mean{c}$, $\Lcal$ is defined to be a linear operator.
Substituting $\eqref{eq:L_def}$ into the mean evolution equation for $\mean{c}$ yields the system
\begin{eqnarray}
    \begin{aligned}
      \diffp{\mean{c}(x,t)}{t} + \mean{u} \diffp{\mean{c}(x,t)}{x} = \nu_p\diffp[2]{\meanc(x,t)}{x} + \Lcal\mean{c}(x,t), \quad x \in (0,L_x), \\
        \mean{c}(0,t) = \mean{c}(L_x,t), \\
        \mean{c}(x,0) = c_{\,0}(x).
    \end{aligned}
    \label{eq:composite_ADE}
\end{eqnarray}
Because $\Lcal$ is linear and defined on a finite domain, it can be specified by its eigenvalues and eigenfunctions, $\lambda_k$ and $f_k$, $k\in\Z$.
Assuming its eigenfunctions form a basis for the solution space of \eqref{eq:composite_ADE}, its action on $\mean{c}$ can be expressed as
\begin{eqnarray}
\begin{aligned}
  \Lcal \mean{c}(x,t) = \sum_{k=-\infty}^{\infty} \lambda_k c_k(t) f_k(x),\\
\end{aligned}
\label{eq:L_eigendecomp}
\end{eqnarray}
where $c_k$ are the expansion coefficients of $\mean{c}$. 
This parametrization enables further constraints to be applied to $\Lcal$ via $\lambda_k$ and $f_k$.

The second constraint is shift invariance, which means $\Lcal$ must commute with the spatial shift operator $\Scal_{x'} f(x) = f(x+x')$ for all $x'$.
The solution space of \eqref{eq:composite_ADE} is the set of continuously-differentiable periodic functions on the bounded domain $[0,L_x]$.
On this domain the shift operator's eigenfunctions are the Fourier modes, $\e{i a_k x},$ where $a_k = 2\pi k / L_x,$ $k\in\Z$.
This implies that the Fourier modes are the eigenfunctions of $\Lcal$ as well, since operators that commute share eigenfunctions.
Let the Fourier coefficients of $\mean{c}$ be denoted $\mean{\chat_k}$ (note that the averaging operator is applied to the coefficients because it commutes with the Fourier transform).
Then the action of $\Lcal$ on $\meanc$ can be expressed in a Fourier series as 
\begin{eqnarray*}
  \begin{aligned}
    \Lcal\mean{c}(x,t) = \sum_{k=-\infty}^{\infty} \lambda_k \mean{\chat_k}(t)\e{ia_k x}, \quad a_k \equiv \frac{2\pi k}{L_x}.
  \end{aligned}
\end{eqnarray*}

Since the eigenfunctions of $\Lcal$ are known, only its eigenvalues $\lambda_k$ are uncertain and are constrained further.
The advection-diffusion equation is a statement of mass conservation, so $\lambda_k$ must be defined so that $\Lcal$'s action does not add mass to the system.
Then
\begin{eqnarray*}
    \begin{aligned}
      0 &= \diff{}{t}\int_0^{L_x} \mean{c}(x,t) \d x \\
      &= \int_0^{L_x} \diffp{\mean{c}(x,t)}{t} \d x \\
      &= \int_0^{L_x} \nu_p \diffp[2]{\mean{c}(x,t)}{x} + \Lcal \mean{c}(x,t) - \mean{u} \diffp{\mean{c}(x,t)}{x}\d x\\
      &= \sum_{k=-\infty}^{\infty}\int_0^{L_x} \left(-\nu_p a_k^2 + \lambda_k - \mean{u}ia_k\right)\mean{\chat_k}(t)e^{ia_k x} \d x\\
        &= \lambda_0 \mean{\chat_0},
    \end{aligned}
\end{eqnarray*}
where all integrals in the sum are zero besides $k=0$ because of periodicity. 
Thus it is sufficient to require $\lambda_0=0$ to conserve mass.

Additional constraints based on expected physical behavior can also be imposed.
For instance, solutions of this physical system are known to decay spatially with time to a uniform $\mean{c}\equiv \mean{\chat_0}$ as the contaminant is diffused throughout the domain.
To ensure this behavior it is sufficient to require $\abs{\mean{\chat_k}}$ decay with time $\forall k\not=0$.
The Fourier coefficients of the solution $\mean{c}$ to \eqref{eq:composite_ADE} are
\begin{eqnarray*}
  \mean{\chat_k}(t)
= \mean{\chat_k}(0)\e{\left(-\nu_pa_k^2 + \lambda_k - \meanu ia_k\right)t \vphantom{\diff{}{x}}}, \quad k \in \Z.
\end{eqnarray*}
Separating this into its real and imaginary parts yields
\begin{eqnarray*}
  \mean{\chat_k}(t) = \mean{\chat_k}(0)\e{\left(-\nu_pa_k^2 + \Re\left[\lambda_k\right]\right)t +i\left(\Im\left[\lambda_k\right] - \meanu a_k\right)t\vphantom{\diffp{}{x}}}, \quad k \in \Z.
\end{eqnarray*}
Only the real part of the argument in the exponential affects the coefficients' magnitude, so
\begin{eqnarray*}
    \begin{aligned}
        \abs{ \mean{\chat_k}(t) } 
        &= \abs{ \mean{\chat_k}(0)}\abs{\;\e{\left(-\nu_p a_k^2 + \Re\left[\lambda_k\right]\right)t} }.
    \end{aligned}
\end{eqnarray*}
Then it is sufficient to require that 
\begin{eqnarray}
  \begin{aligned}
    -\nu_p a_k^2 + \Re\left[\lambda_k\right] \leq 0
  \end{aligned}
\end{eqnarray}
to guarantee the solution fluctuations will not grow with time. 

An additional property of this system is that the mean concentration should be propagated downstream. 
The imaginary part of the operator affects the advection of $\meanc$, which can be seen by rearranging the evolution equation of $\mean{\chat_k}$:
\begin{eqnarray*}
  \begin{aligned}
    \diff{\mean{\chat_k}}{t} + i\Big( \mean{u} a_k - \Im[\lambda_k]\Big)\mean{\chat_k} 
    = \Big(-\nu_p a_k^2 + \Re[\lambda_k]\Big)\mean{\chat_k}.
  \end{aligned}
\end{eqnarray*}
The velocity at which Fourier mode $k$ propagates is $\mean{u}-\Im[\lambda_k]/a_k$, so to guarantee downstream propagation for all wavenumbers it is sufficient to require
\begin{eqnarray}
  \begin{aligned}
    \mean{u}a_k - \Im[\lambda_k] > 0.
  \end{aligned}
\end{eqnarray}

To this point the physical constraints placed on the operator resulted in simple constraints on its structure that were easy to impose. 
A further constraint on $\Lcal$ is that $\meanc$ must remain positive. 
Determining a constructive constraint to enforce this property is challenging. 
The typical approach of reformulating the problem in terms of the $\log \meanc$ makes the governing equations nonlinear, which precludes the use of the eigenfunction expansion of $\meanc$ to parametrize the action of $\Lcal$. 
Conditions on the Fourier expansion of a function to guarantee positivity, a likely means of deriving such constraints for $\Lcal$, are an open area of inquiry.
Due to these challenges a positivity constraint was not enforced here.

Note that $\Lcal$ should exactly represent the effects of dispersion on the evolution of $\mean{c}$, since, by definition,
\begin{eqnarray*}
  \begin{aligned}
    \Lcal\mean{c}(x,t) = - \diffp{\mean{u'c'}(x,t)}{x}.
  \end{aligned}
\end{eqnarray*}
In terms of the Fourier series solution of $\mean{c}$, this equates to $\lambda_k \mean{\chat_k}(t) = - (ia_k)  \mean{\widehat{\left(u'c'\right)}_k}(t)$.
Solving for $\lambda_k$ yields
\begin{eqnarray*}
  \begin{aligned}
    \lambda_k &= \frac{-(ia_k)\mean{\widehat{\left(u'c'\right)}_k}(t)}{\mean{\chat_k}(t)}.
  \end{aligned}
\end{eqnarray*}
This highlights that, unless $\mean{(u'c')_k}(t)/\mean{\chat_k}(t)$ is a constant proportion as a function of time, an exact representation would in general require time dependence in $\lambda_k$.
This time dependence cannot be recovered directly, however, because $\mean{u'c'}$ cannot be observed.
For this initial study, $\lambda_k$ are assumed constant in time for simplicity. 
It should be noted that this assumption induces inadequacy in the formulation, which will limit its ability to successfully extrapolate in time.
This completes the deterministic formulation of $\Lcal$ used in this work.

\section{Bayesian inference problem specification}\label{sec:ip_formulation}
In \Cref{sec:operator_formulation}, the uncertain operator $\Lcal$ was parametrized by its eigenvalues and eigenfunctions, and its eigenfunctions were determined to be the Fourier modes.
The only remaining uncertainty in $\Lcal$ is in its eigenvalues $\lambda_k$, for which a Bayesian inverse problem will be defined. 
Observations of the mean concentration $\mean{c}$ at different times and locations constitute the data for the inverse problem.
How many eigenvalues can be inferred, and how precisely, is assessed as a function of the frequency of observation and as a function of whether observations are collected in a time series or across the spatial domain.
Samples of the posterior distribution are generated using Markov Chain Monte Carlo (MCMC).

Let the right-hand side of \eqref{eq:composite_ADE} be denoted $\Dcal$ so that 
\begin{eqnarray}
    \mathcal{D}\mean{c} \equiv \left(\nu_p\diffp[2]{}{x} + \Lcal\right)\mean{c}.
    \label{eq:D}
\end{eqnarray}
The uncertainty of $\Lcal$ induces uncertainty in $\Dcal$, whose eigenvalues are denoted $\mu_k$. 
As discussed in \Cref{sec:operator_formulation}, the eigenfunctions of $\Dcal$ are the Fourier modes because of the shift invariance of $\Lcal$ and the second derivative. 
The eigenvalues of $\Dcal$ are therefore given by 
\begin{eqnarray}
	\mu_k = -\nu_p a_k^2 + \lambda_k,
    \label{eqn:muLambda}
\end{eqnarray}
so that inference of $\mu_k$ is equivalent to inference of $\lambda_k$.

Inference will be formulated in terms of $\mu_k$ for two reasons. 
First, it is simpler to enforce the constraint $\Re\left[ \mu_k \right] \leq 0$ to guarantee that $\abs{\mean{\chat_k}}$ decay with time, as discussed in \Cref{sec:operator_formulation}.
Second, it will allow for direct comparison with another popular model of contaminant transport through heterogeneous media, the fractional advection-diffusion equation (FRADE) \cite{berkowitz2006ctrw,schumer2009fractional}.
Given the parametrization of the uncertain operator using its eigendecomposition \eqref{eq:L_eigendecomp}, the goal is to infer the uncertain eigenvalues of $\Dcal$, which forms the right-hand side of a generalized diffusion equation:
\begin{eqnarray}
\begin{aligned}
  \diffp{\mean{c}(x,t)}{t} + \mean{u} \diffp{\mean{c}(x,t)}{x} = \Dcal\mean{c}(x,t), \\
        \mean{c}(0,t) = \mean{c}(L_x,t), \\
        \mean{c}(x,0) = c_0(x).
    \end{aligned}
\label{eq:generalized_ADE}
\end{eqnarray}

%
\subsection{Prior specification}
Prior distributions are defined on the real and imaginary parts of $\mu_k$, and their posterior distributions are inferred using observations of $\mean{c}$ in a Bayesian inverse problem.
Because $\mean{c}$ is real, its Fourier coefficients are conjugate symmetric; that is, $\mean{\chat_{-k}} = \overline{\mean{\chat_k}}$, where $\overline{(\cdot)}$ represents the complex conjugate.
As a result, the action of $\Dcal$ on $\mean{c}$ can be expressed using just the eigenvalues associated with positive wavenumbers:
\begin{eqnarray*}
  \begin{aligned}
     \Dcal\mean{c}(x,t) =2 \,\Re\left[\sum_{k=1}^{\infty} \mean{\chat_k}(t) \mu_k \e{ia_k x} \right],
  \end{aligned}
\end{eqnarray*}
where the series is indexed from $1$ because $\mu_0=0$ to conserve mass.
An upper bound on the number of eigenvalues to be inferred is $N_k$,
the number of terms in a similar Fourier expansion needed to resolve the initial condition $c_0(x)$. 
This is because the Fourier coefficients of $\meanc$ decay with increasing $k$ and time, so the number of terms needed to resolve $\meanc$ will never exceed those required to resolve $c_0$. 
It should be noted that for problems that exhibit this property, the dimensionality of the inverse problem for the operator is determined by the spectral content of the solution, not by the discretization of the problem, as it would be if the discretized values of the operator were inferred directly.

The prior distributions for the real and imaginary parts of the eigenvalues of $\Dcal$ are defined using the deterministic constraints discussed in \Cref{sec:operator_formulation} and considerations of what values they would plausibly take.
For the real part of the eigenvalues, recall that $\Re[\mu_k] \leq 0$ to ensure $\abs{\mean{\chat_k}}$ decay with time. 
This is a hard upper bound for the real parts that cannot be violated.
A plausible lower bound for the real parts is determined by observing that the value of an eigenvalue cannot be inferred from the data if the corresponding Fourier mode is too rapidly damped.
The plausible lower bound is therefore set as the eigenvalues of a diffusion operator with diffusion coefficient $\nu_{max}$ that ensures that at least the lowest wavenumber Fourier coefficient does not decay by more than some factor $A$ in magnitude while propagating the length of the domain $L_x$ at the velocity $\meanu$.
In this case, the value of $\nu_{max}$ is given by
\begin{equation}
  \nu_{max}=\frac{L_x\meanu\ln A}{4\pi^2}.
\end{equation}
For the values of $L_x$ and $\meanu$ used in \Cref{sec:case_1,sec:case_2} ($L_x=4$, $\meanu=1$) and $A=10^{10}$ one obtains $\nu_{max}\approx 2.5$, which is used to define the priors below.
The posteriors determined in \Cref{sec:case_1,sec:case_2} are dominated by the likelihood, so the plausible lower bounds defined here have no impact on the inference.

The imaginary parts of $\mu_k$ are identical to those of $\lambda_k$, so they must satisfy the same condition to guarantee downstream propagation for this problem:
\begin{eqnarray*}
  \begin{aligned}
    \mean{u}a_k - \Im[\mu_k] > 0 \implies \Im[\mu_k] < \mean{u}a_k.
  \end{aligned}
\end{eqnarray*}
This is a hard upper bound on the imaginary parts of the eigenvalues. 
A plausible lower bound of $\Im[\mu_k] \geq - \mean{u}a_k$ is imposed. 
This bound implies the contribution to transport from dispersion will be less than or equal to the contribution from advection, which is expected in practice.
As with the plausible bound for the real parts of the eigenvalues, the prior is dominated by the likelihood in the posterior, so this specification was determined not to have an effect on the results.

For both the real and imaginary parts of the eigenvalues a hard bound is specified on one side of the domain, while a plausible bound has been placed on the other.
The hard bound should not be violated, while there is no physical reason that an eigenvalue cannot go outside the plausible bounds that have been set, though exceeding the bounds is considered improbable. 
To reflect this state of knowledge, the prior distributions for the real and imaginary parts will be defined as exponential distributions, with the hard bounds corresponding to the lower bound of the exponential distribution and the plausibility bounds used to define the scaling coefficients.

Let $\Re[\mu_k]\equiv R_k$. 
The bounds on $R_k$ are $-\nu_{max} a_k^2 \lesssim R_k < 0,$ where the $\lesssim$ denotes the plausible bound.
The negative real parts $-R_k$ are then represented using the exponential distribution: 
\begin{eqnarray*}
  \begin{aligned}
    p(-R_k ) = \e{-(-R_k)/\beta_k}/\beta_k.
  \end{aligned}
\end{eqnarray*}
The scaling coefficients $\beta_k$ are defined so that $95\%$ of the probability mass for each $-R_k$ falls between 0 and $\nu_{max} a_k^2$.
This is done using the CDF of $-R_k$, $P(-R_k) = 1 - \e{-(-R_k)/\beta_k}$:
\begin{eqnarray*}
  \begin{aligned}
    0.95 &=  P(-(-\nu_{max} a_k^2)) = 1-\e{-\frac{\nu_{max} a_k^2}{\beta_k}}  \\
    &\Downarrow\\
    \beta_k &= -\nu_{max} a_k^2 / \ln{0.05}.
  \end{aligned}
\end{eqnarray*}

\begin{figure}[h]
    \centering
    \includegraphics[scale=.8]{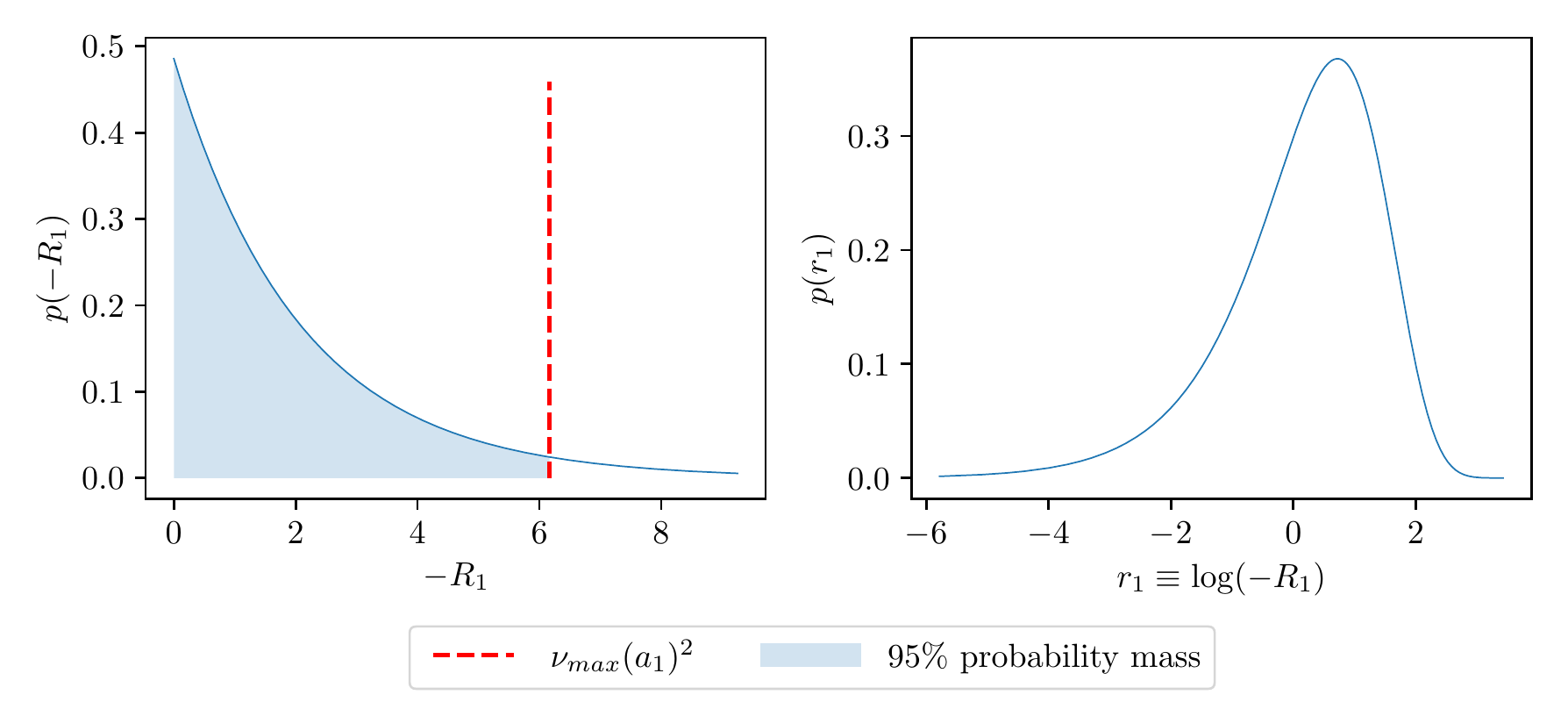}
    \caption{The prior distribution for $-R_1$ and $r_1$.}
    \label{fig:real_prior}
  \end{figure}

The negative real parts are bounded from below by zero. 
This can harm mixing for MCMC algorithms that employ Gaussian proposal distributions, which can generate many samples outside the parameter domain.
To avoid the bound, the transformed variables $r_k=\log(-R_k)$ were inferred instead.
Their prior distributions can be computed analytically using a variable transformation and are defined as
\begin{eqnarray}
  \begin{aligned}
    r_k = \log(-R_k), \\
    p(r_k) = \frac{\e{-e^{r_k}/\beta_k + r_k}}{\beta_k,},
  \end{aligned}
  \label{eq:real_part_prior}
\end{eqnarray}
where $\beta_k$ are the same as for the distributions of $R_k$.
This prior distribution and the distribution of $-R_1$ is shown in \Cref{fig:real_prior} for reference.

Let $\Im[\mu_k]\equiv I_k.$
The bounds on $I_k$ are $ -\mean{u}a_k \lesssim I_k < \mean{u} a_k$, where again $\lesssim$ denotes the plausible bound.
To derive a variable that is consistent with the exponential distribution, let $U_k = \mean{u}a_k - I_k$.
Then $ 0 < U_k \leq 2\mean{u}a_k$, and the scaling coefficients of the exponential distribution are defined so that 95\% of the probability mass falls between these bounds as was done for $-R_k$.
To avoid the hard lower bound on $U_k$ the transformed variables $u_k=\log(U_k)$ were inferred instead. 

Inference of the  parameters $\Thetavec\equiv[r_1,r_2, \cdots, u_1, u_2,\cdots]$ is performed. 
The real and imaginary parts of the eigenvalues are then recovered using the transformations
\begin{eqnarray}
  \begin{aligned}
    R_k &= - e^{r_k},  \\
    I_k &= \mean{u}a_k - e^{u_k}.
  \end{aligned}\label{eq:parameter_mappings}
\end{eqnarray}
All parameters are assumed independent, so the infinite-dimensional prior density is defined as 
\begin{eqnarray*}
  p(\thetavec)\equiv \prod_{k=1}^{\infty} p(r_k)p(u_k).
\end{eqnarray*}
Recall, however, that the number of inferred eigenvalues is limited to at most $N_k$, where $N_k$ is the number of Fourier modes required to resolve the Fourier expansion of the initial condition.
The truncated version of the prior is thus defined as
\begin{eqnarray}
  p(\thetavec)\equiv \prod_{k=1}^{K} p(r_k)p(u_k),
  \label{prior}
\end{eqnarray}
where $K\leq N_k$.
$K$ is determined by a global sensitivity analysis, described in \Cref{sec:inference_implementation}.

\subsection{Likelihood specification}
Several data sets will be considered, but in all cases an additive, normally-distributed measurement error is assumed.
The data model is denoted
\begin{eqnarray*}
    \begin{aligned}
      d_{i} = \mean{c}(x_i, t_i; \Thetavec) + \eps_i, \quad i = 1, \cdots, N_{obs}, \quad \eps_i \sim \mathcal{N}( 0, \Sigma ),
    \end{aligned}
\end{eqnarray*}
where $\Sigma$ is the measurement error covariance matrix.
Then the likelihood is defined as
\begin{equation}
  \begin{aligned}
    p(\dvec | \Thetavec ) & = \frac{
      \e{ -\frac{1}{2} \norm{ \dvec - \mean{\cvec}}_{\Sigma^{-1/2}}^2 }
      }{
        (2\pi)^{N_{obs} / 2}|\Sigma|^{-1/2}
      }.
  \end{aligned}
  \label{eq:likelihood}
\end{equation}

\subsection{Inference implementation}\label{sec:inference_implementation}
Before inference a global variance-based sensitivity analysis is performed to determine to which eigenvalues $\mean{c}$ is most sensitive \cite{gbvsa1,gbvsa2,gbvsa3}.
The sensitivity is assessed by computing the Sobol' total-effect index, a measure of the contribution to variance in $\mean{c}$ from varying an eigenvalue alone as well as from its variation along with other eigenvalues.  
Any eigenvalues whose Sobol total-effect indices exceed a heuristic threshold are included in the inference, and the rest are fixed at a reasonable value as described below.
This heuristic threshold was determined by studying mixing of Markov chains at different threshold values and selecting the value that produced the best mixing.
The sensitivity analysis is performed using the Python software package SALib \cite{salib}.

To generate samples of the posterior distributions of the eigenvalues using Markov Chain Monte Carlo (MCMC), the Delayed Rejection Adaptive Metropolis (DRAM) algorithm \cite{DRAM}, implemented in Version 1 of the MIT UQ Library (MUQ1) \cite{MUQ}, is used.
The starting point of the Markov chain is determined by performing two deterministic optimizations, also using MUQ.
The first optimization is performed with the assumption that $\Dcal$ is of the form
\begin{eqnarray*}
\begin{aligned}
\Dcal = \nu \diffp[\alpha]{}{x},
\end{aligned}
\end{eqnarray*}
and $\nu$, $\alpha$ are optimized to maximize the likelihood density.
The second optimization is initialized at the solution to the first optimization and relaxes the assumption on the form of $\Dcal$, optimizing over $r_k$ and $u_k$ to maximize the posterior.
Only the eigenvalues to which $\mean{c}$ is sensitive are optimized and included in the Bayesian inference.
To ensure the uninferred eigenvalues are set at reasonable values, they are fixed at the solution of the first optimization.
By fixing the insensitive eigenvalues at those of the fractional derivative, the Bayesian inference can be interpreted as finding a correction to the fractional derivative for the sensitive eigenvalues.

\subsection{Analysis of MCMC results}
Chains of length $3\times 10^5$ were run, and the first $1\times10^5$ samples were discarded as burn-in for each of the cases discussed. 
The Kullback-Leibler divergence (DKL or KL divergence) is a natural measure of how much information was gained through inference \cite{kl_div}, since it is a measure of how different two probability distributions are from each other.
Of interest for this work is the KL divergence between the marginal prior and posterior for each parameter, since this will provide an assessment of how much information is gained about the eigenvalues as a function of $k$.
The DKL between prior and posterior for a single parameter $\Theta_k$ is defined as
\begin{eqnarray*}
\begin{aligned}
  D\Big( p(\Theta_k|\dvec) \;\Big|\Big|\; p(\Theta_k ) \Big) &= \int \ln\left( \frac{p(\Theta_k|\dvec)}{p(\Theta_k)} \right) p(\Theta_k | \dvec) \d \Theta_k.
\end{aligned}
\end{eqnarray*}
Larger values of KL divergence indicate greater information gain between posterior and prior.

If an analytical expression for the marginal posterior were available, this integral could be approximated using Monte Carlo integration by
\begin{eqnarray*}
\begin{aligned}
  D\Big( p\left(\Theta_k|\dvec\right) \;\Big|\Big|\; p\left(\Theta_k\right) \Big) \approx \frac{1}{N_s}\sum_i^{N_s} \ln\left[p\left(\Theta_k^{(i)}\Big|\dvec\right)\right] - \ln\left[p\left(\Theta_k^{(i)}\right)\right], \; \Theta_k^{(i)} \sim p\left(\Theta_k|\dvec\right),
\end{aligned}
\end{eqnarray*}
where $N_s$ is the number of samples used in the sample mean.
Because an analytical expression is unavailable, an approximation of the posterior distribution must be provided. 
It is common to approximate the posterior using a Kernel-Density Estimate (KDE) approximation \cite{kde}, built using samples from the posterior generated using MCMC.
However, in this case, MCMC samples of the posterior indicate a
nearly-Gaussian posterior
so a Gaussian approximation was used.
The KL divergence between a single parameter $\Theta_i$'s marginal posterior and prior is thus approximated by
\begin{eqnarray}
\begin{aligned}
  D\Big(p(\Theta_k|\dvec) \,\Big|\Big|\, p(\Theta_i) \Big) \approx \frac{1}{N_s}\sum_{i=1}^{N_s} \ln\left[ p_{GA}\left(\left.\Theta_k^{(i)} \,\right|\, \dvec\right) \right] - \ln\left[ p\left(\Theta_k^{(i)}\right)\right], \; \Theta_k^{(i)}\sim p(\Theta_k | \dvec ),
\end{aligned}
\label{eq:DKL}
\end{eqnarray}
where $p_{GA}$ denotes a Gaussian approximation.

\section{Case 1: Data from Fractional Advection-Diffusion Equation}\label{sec:case_1}
\begin{figure}[h]
    \centering
    \includegraphics[scale=0.8]{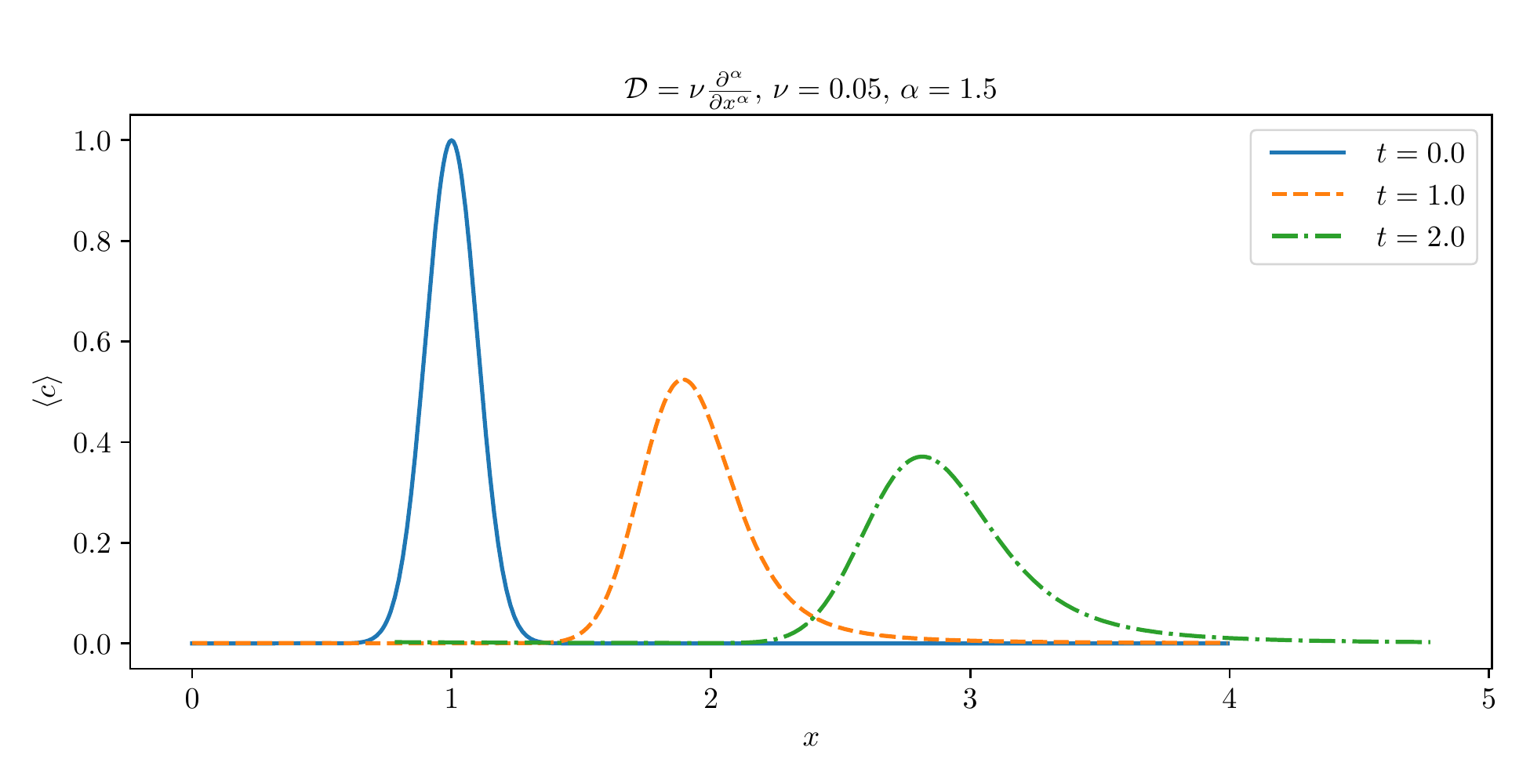}
    \caption{The evolution of a Gaussian initial condition with the FRADE at $t=1$ and $t=2$, or $\sfrac{1}{4}$ and $\sfrac{1}{2}$ of a flowthrough time, respectively.}
    \label{fig:FRADEevolution}
\end{figure}

To study how successful Bayesian inference can be in the case where $\Dcal$ could exactly represent the underlying operator, data was generated using a 1D fractional advection-diffusion equation (FRADE), defined as
\begin{equation}
	\begin{aligned}
    \diffp{\meanc(x,t)}{t} + \meanu \diffp{\meanc(x,t)}{x} = \nu \diffp[\alpha]{\meanc(x,t)}{x},& \quad x\in (0,4), \quad \alpha \in [1,2] \\
		\meanc(0, t) = \meanc(4, t), &\\
                \meanc(x,0) = \e{ -\frac{(1-x)^2}{2 \beta^2} }&\qquad\mbox{with $\beta=0.1$} .
	\end{aligned}
	\label{eq:FRADE}
\end{equation}
  The data used for inference in this section was generated with $\meanu=1$, $\alpha=1.5$, and $\nu=0.05$ and taken at times $(x_i,t_i)$ from a Fourier series solution of \eqref{eq:FRADE} on a 512-point regular spatial grid, with Fourier coefficients defined as 
\begin{eqnarray*}
  \mean{\chat_k}(t_i) = \mean{\chat_k}(0) \e{ \left[ \nu(ia_k)^\alpha - \meanu (ia_k) \right]t_i }.
\end{eqnarray*}
Fractional PDEs can be seen as limiting forms to solutions of continuous-time-random-walk models, which are popular representations of anomalous diffusion in heterogeneous porous media \cite{berkowitz2006ctrw}.
An example of the time evolution of the concentration field generated from this model is shown in Figure \ref{fig:FRADEevolution}.
In this case it is known \textit{a priori} that the true eigenvalues of $\Dcal$ are $\mu_k = \nu (ia_k)^\alpha$, making it possible to study if the true values are recovered in different data scenarios.

\subsection{Likelihood}
Data was generated by sampling the FRADE solution over a range of times and locations. 
Random noise distributed according to $\Ncal(0,\sigma^2)$  was added to the model evaluations to simulate measurement error.
The measurement standard deviation of $\sigma=0.005$ corresponding to a $1\%$ standard error in the maximum concentration $\mean{c}=1$ was used, making the form of the likelihood
\begin{eqnarray*}
  \begin{aligned}
    p(\dvec | \Thetavec ) = \frac{1}{(2\pi \sigma^2)^{N/2}}\e{ -\frac{1}{2\sigma^2}\norm{ \dvec - \mean{\cvec} }_2^2 }, \quad \sigma=0.005.
  \end{aligned}
\end{eqnarray*}

\subsection{Results}\label{sec:case1_results}
The eigenvalues of $\mathcal{D}$ were inferred using spatial- and time-series data with 32, 64, or 512 observations, taken at regular intervals across the entire spatial domain, 
or in time from $t=0$ to $4$, which is the time required to advect the length of the domain at velocity $\mean{u}=1$. 
In both cases, the entirety of the pulse and its tails was observed.


Global variance-based sensitivity analysis was performed to determine how many eigenvalues to infer for each data scenario.
The number of eigenvalues whose Sobol indices exceeded the tolerance of $10^{-4}$ for each scenario are reported in \Cref{tab:sensitive_eigenvalues}.
The number of sensitive eigenvalues did not depend on the number of observations taken over the range investigated (from 32 to 512), so only the observation location or time for time-series or spatial-series data is reported.
The decay of higher wavenumber coefficients in the Fourier series solution of $\mean{c}$ as a function of time mean that an upper bound on the number of possibly sensitive eigenvalues is set by the number of modes initially excited in the initial condition.
  For the initial condition specified in \Cref{eq:FRADE}, the first 47 modes were excited, based on a threshold of $10^{-13}$ for the modulus of the coefficient.
Additionally, the decay of coefficients with time means that the number of modes that remain excited and thus the eigenvalues to which $\mean{c}$ is sensitive will decay accordingly.
This is demonstrated by the spatial-data cases, where the number of sensitive eigenvalues decreased with increasing time.
Cases with spatial data observed at a single time were sensitive to more eigenvalues than cases with time-series data observed at a single location.
The maximum number of eigenvalues that were informed over all the cases considered was 10, significantly fewer than the number of initially excited modes in the initial condition.

\begin{table}\centering
\begin{tabular}{@{}cccc@{}}\toprule
\multicolumn{2}{c}{\textbf{Spatial series}} & \multicolumn{2}{c}{\textbf{Time series}}  \\
Observation time & \# sensitive eigenvalues & Observation location & \# sensitive eigenvalues \\\hline
0.5 & 10 & 2.0 & 5 \\
1.0 & 8 & 3.0 & 5 \\
2.0 & 7 & 4.0 & 5 \\
\bottomrule
\end{tabular}
\caption{The number of sensitive eigenvalues for different data
  scenarios, with data drawn from solutions of the FRADE (\ref{eq:FRADE})
  with $\alpha=1.5$ and $\nu=0.05$. Observations are equally spaced in
  space or time, over the whole spatial domain or over $t\in[0,4]$. }
\label{tab:sensitive_eigenvalues}
\end{table}

\begin{figure}[h]
  \centering
  \includegraphics[scale=0.8]{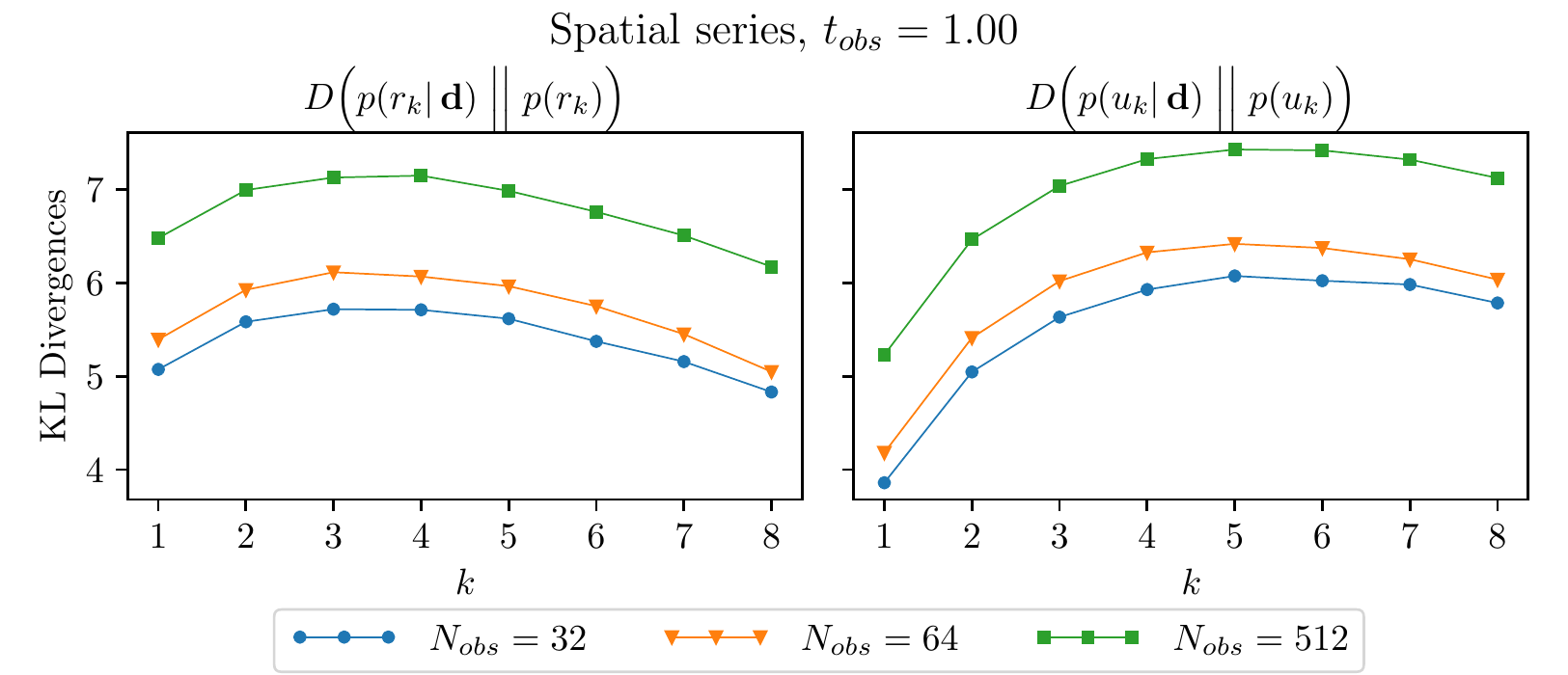}
  \caption{KL divergences of posteriors relative to priors for Bayesian
    inference from spatial-series data with varying number of
    observation $N_{obs}$ evenly spaced through the spatial domain of
    the solution of (\ref{eq:FRADE}).}
  \label{fig:kl_div_t_1}
\end{figure}

\begin{figure}[h]
  \centering
  \includegraphics[scale=.8]{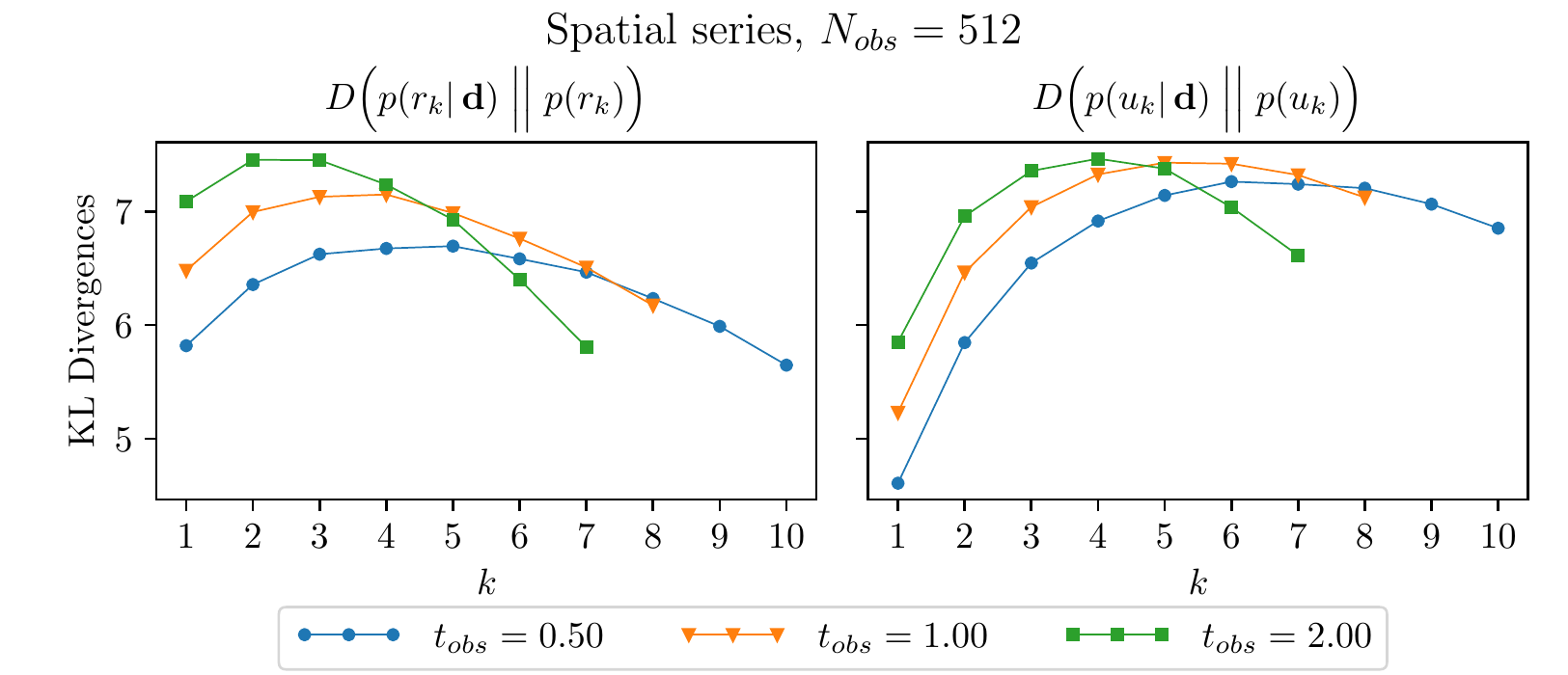}
  \caption{KL divergences of posteriors relative to priors for Bayesian
    inference from spatial-series data with varying observation time
    $t_{obs}$ of the solution of (\ref{eq:FRADE}).}
  \label{fig:spatial_kl_div_nobs_512}
\end{figure}

KL divergences of the posterior relative to the prior for inference using spatial-series data are presented in \Cref{fig:kl_div_t_1} and \Cref{fig:spatial_kl_div_nobs_512}.
In the case of \Cref{fig:kl_div_t_1}, the frequency of observation was varied.
In the case of \Cref{fig:spatial_kl_div_nobs_512}, the time at which the data was collected was varied.
As shown in \Cref{fig:kl_div_t_1}, increased frequency of observation in the spatial domain led to higher information gain in the eigenvalues that were informed by the data, as indicated by a higher KL divergence.
The number of eigenvalues that were informed by the data depended on the time at which the spatial observations were made, as seen in \Cref{fig:spatial_kl_div_nobs_512}.
For successively later times, the solution was sensitive to fewer and fewer eigenvalues.
  It is interesting to note that the information gain for the lowest-frequency eigenvalues increases with later observation times, but that the decay in the KL divergence as a function of $k$ was more rapid for later observations.
  The more rapid decay for later observations is not surprising, since the timescales on which the Fourier modes are dampened varies inversely with their wavenumber. 
  The more rapid damping of high wavenumber modes makes them more difficult to observe at later times.
Conversely, the lowest wavenumbers evolve slowly, which makes their evolution more difficult to observe at early times. 
This is presumably why the KL divergence for the lowest wavenumber modes increases with observation time. 

It is infeasible in realistic applications to have abundant spatial observations of the concentration field, since obtaining data from each location would require the creation of a different well.
Instead it is more likely that one would have access to time-series observations of concentration at a limited number of locations.
To reflect this, time-series data at one location was also used in the Bayesian inference of the eigenvalues.

\begin{figure}[h]
  \centering
  \includegraphics[scale=0.8]{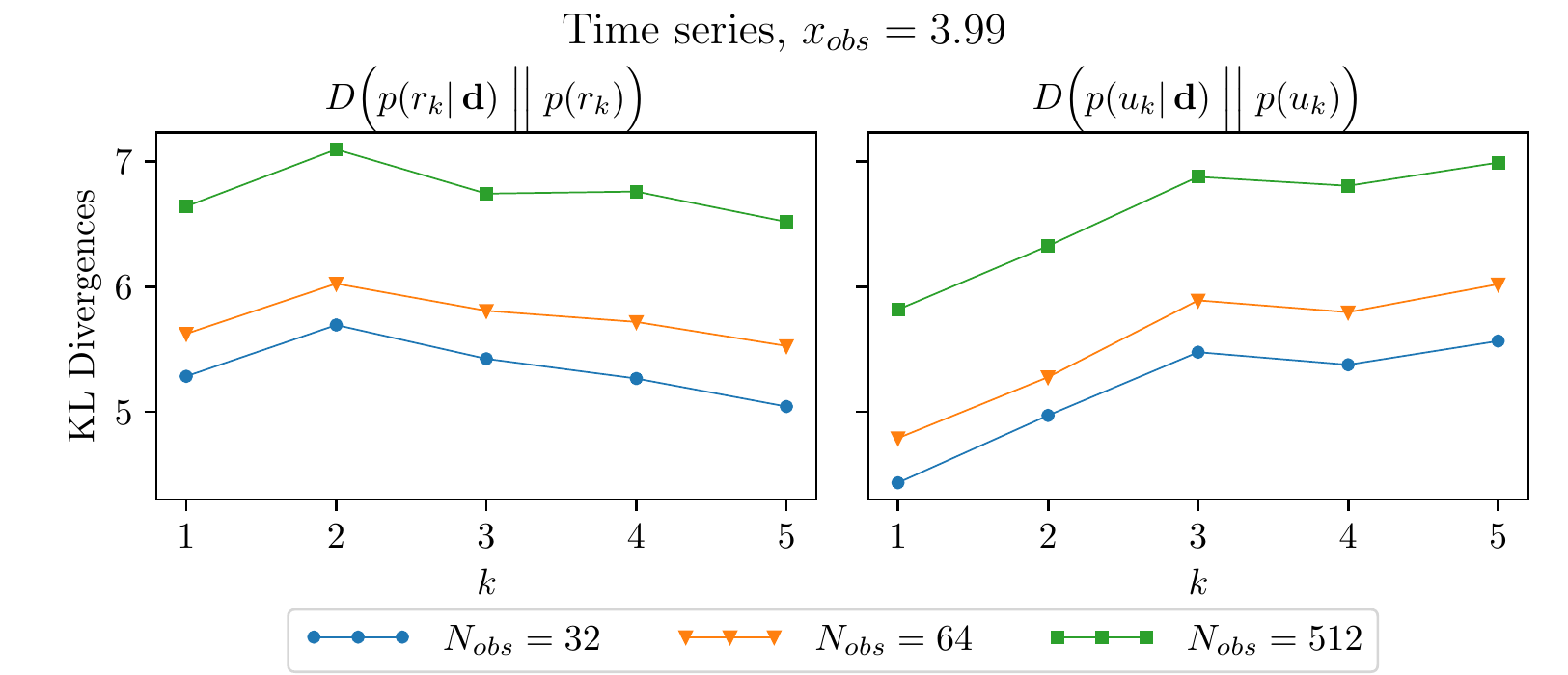}
  \caption{KL divergences of posteriors relative to priors for Bayesian
    inference from time series data of varying number of observations $N_{obs}$ of 
    the solution of (\ref{eq:FRADE}) evenly spaced in the time period $[0,4]$.}
  \label{fig:kl_div_x_4.0}
\end{figure}

\begin{figure}[h]
  \centering
  \includegraphics[scale=0.8]{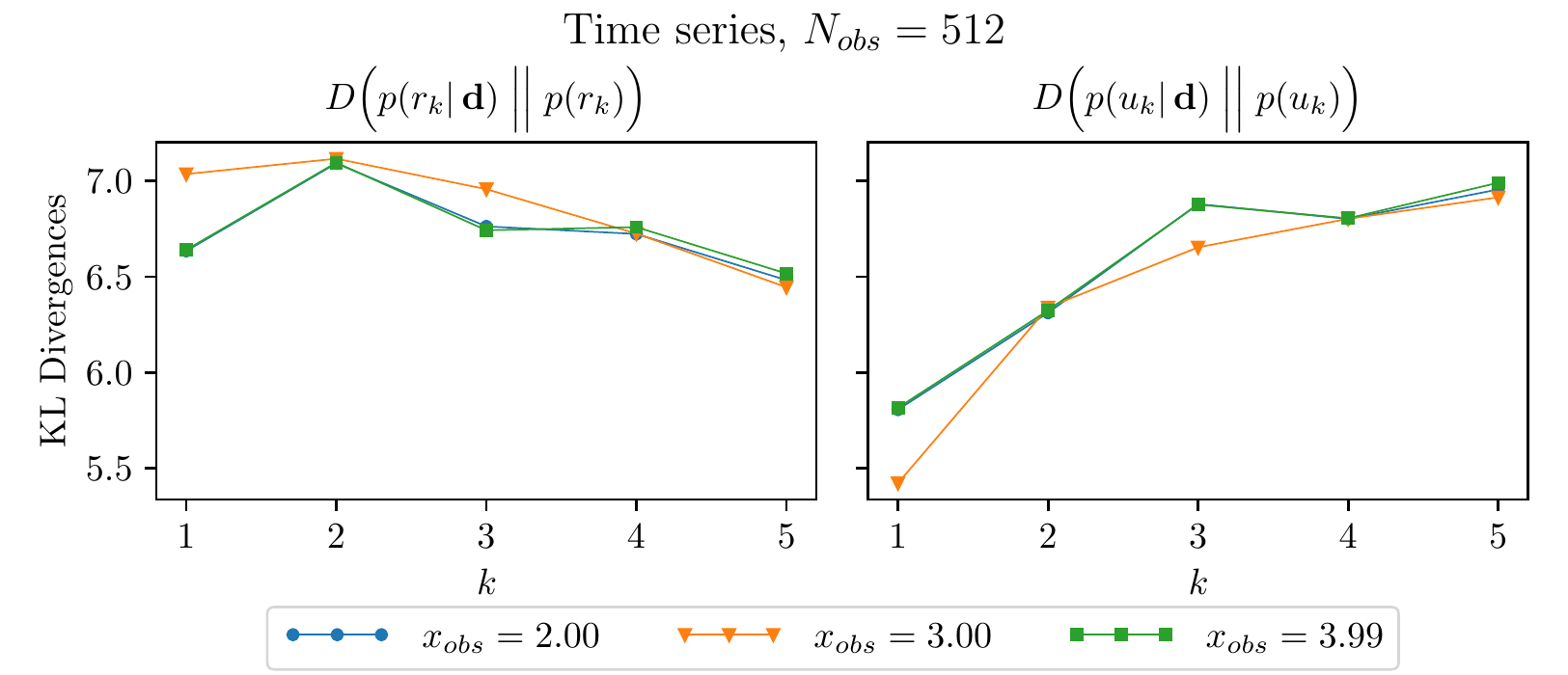}
  \caption{KL divergences of posteriors relative to priors for Bayesian
    inference from time-series data from varying observation locations $x_{obs}$ of 
    the solution of (\ref{eq:FRADE}).}
  \label{fig:time_kl_div_nobs_512}
\end{figure}

\begin{figure}[h]
  \centering
  \includegraphics[scale=0.8]{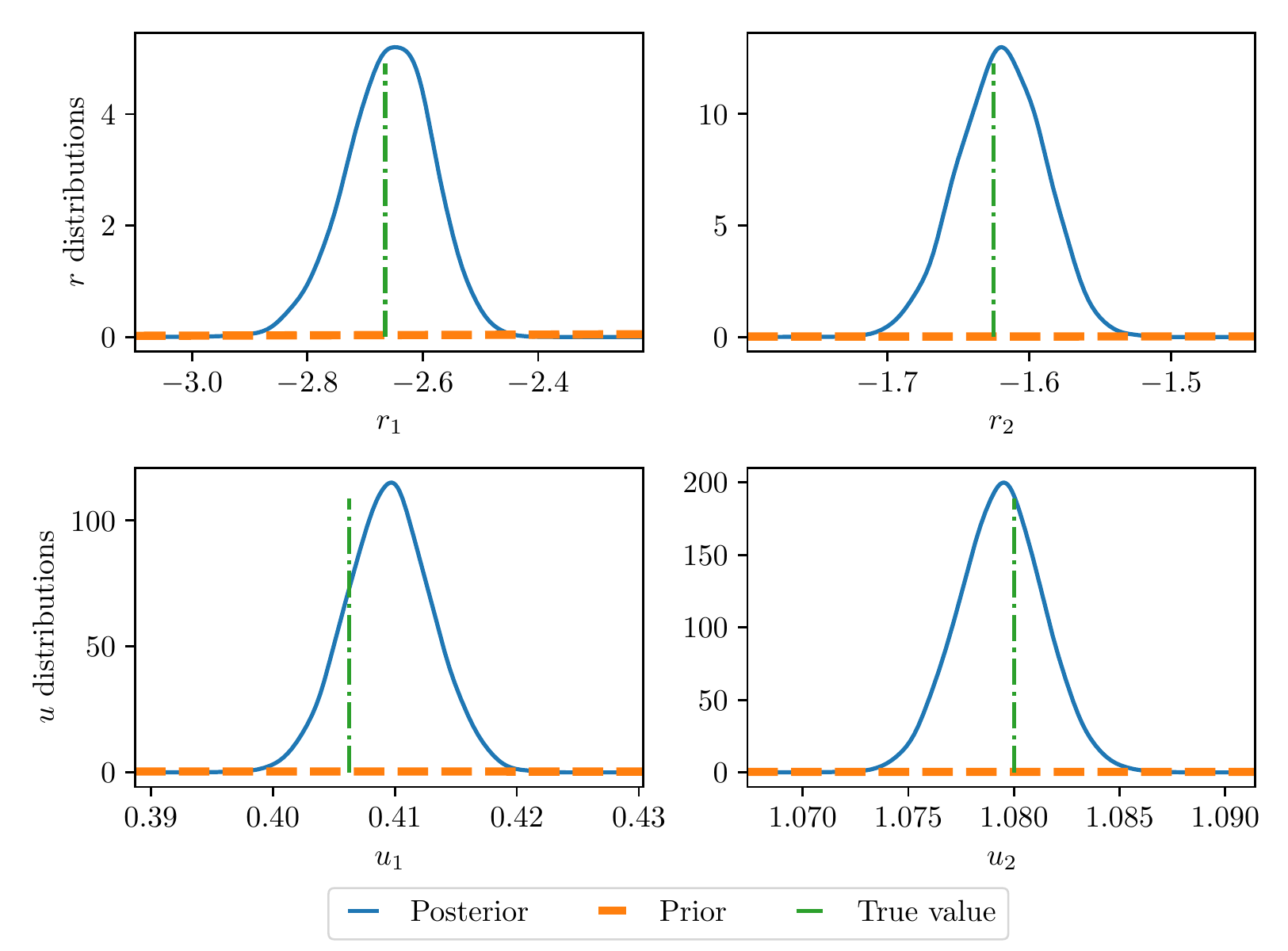}
  \caption{The posterior and prior marginal probabilities for the real
    and imaginary parts of the first two eigenvalues, inferred from
    512 spatial observations of the solution of \eqref{eq:FRADE} equally spaced over the spatial domain, taken at time $t=0.5$.
  \label{fig:case1_n512_posteriors}
}
\end{figure}

\begin{figure}[h]
  \centering
  \includegraphics[scale=.9]{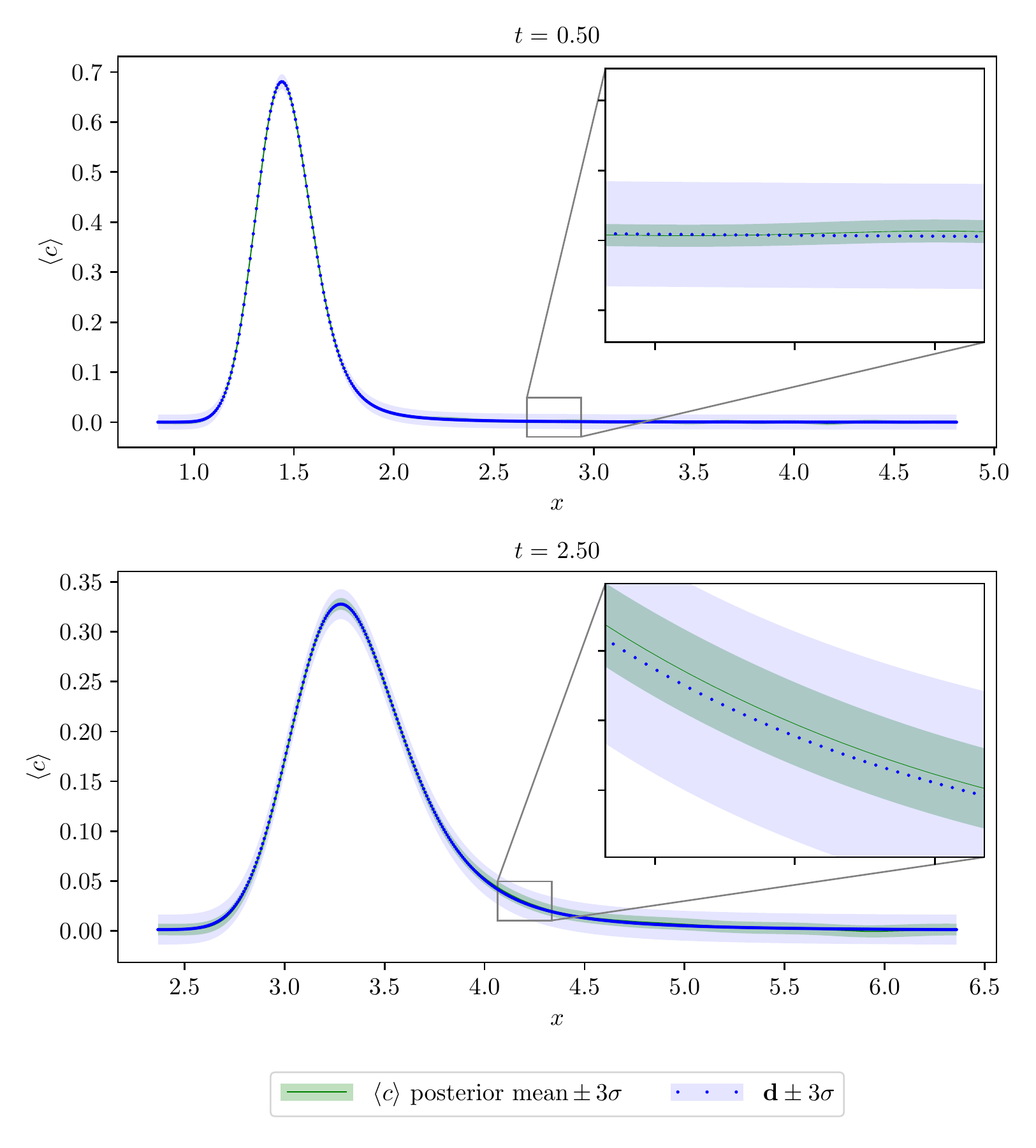}
  \caption{
    Statistics of $\mean{c}$ evolved using posterior samples from inference
  with 512 equally-spaced spatial observations of the solution of
    (\ref{eq:FRADE}) collected at $t=0.5$.
  }
  \label{fig:extrapolated_spatial_stats}
\end{figure}

\begin{figure}[h]
  \centering
  \includegraphics[scale=.8]{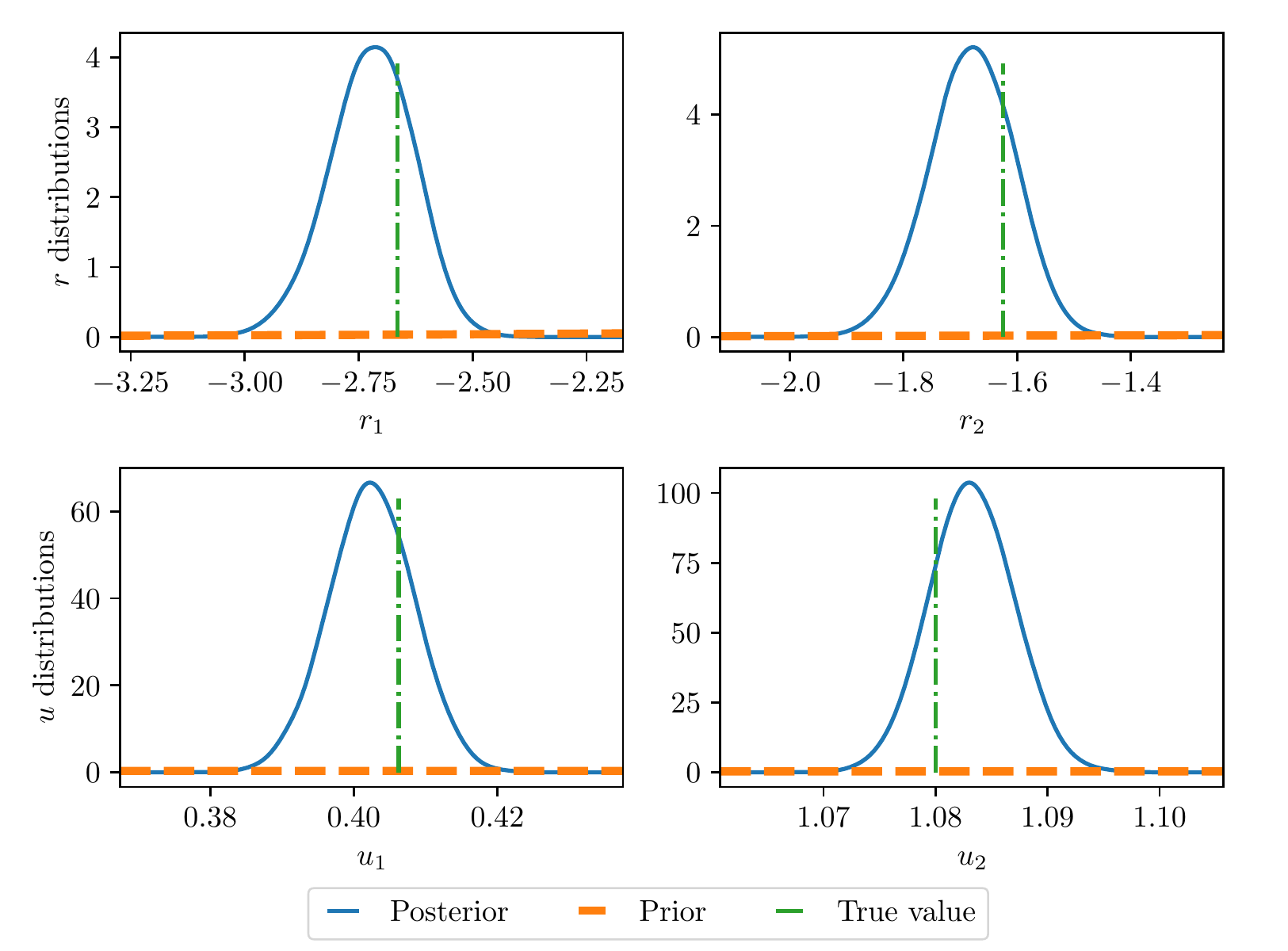}
  \caption{The posterior and prior marginal probability densities for
    the real and imaginary parts of the first two eigenvalues of $\Dcal$,
    inferred using 32 time-series observations of the solution of (\ref{eq:FRADE})
    taken at $x=2.0$ and uniformly spaced over $[0,4]$ in time.}
  \label{fig:case1_n32_posteriors}
\end{figure}

\begin{figure}[h]
  \centering
  \includegraphics[scale=.9]{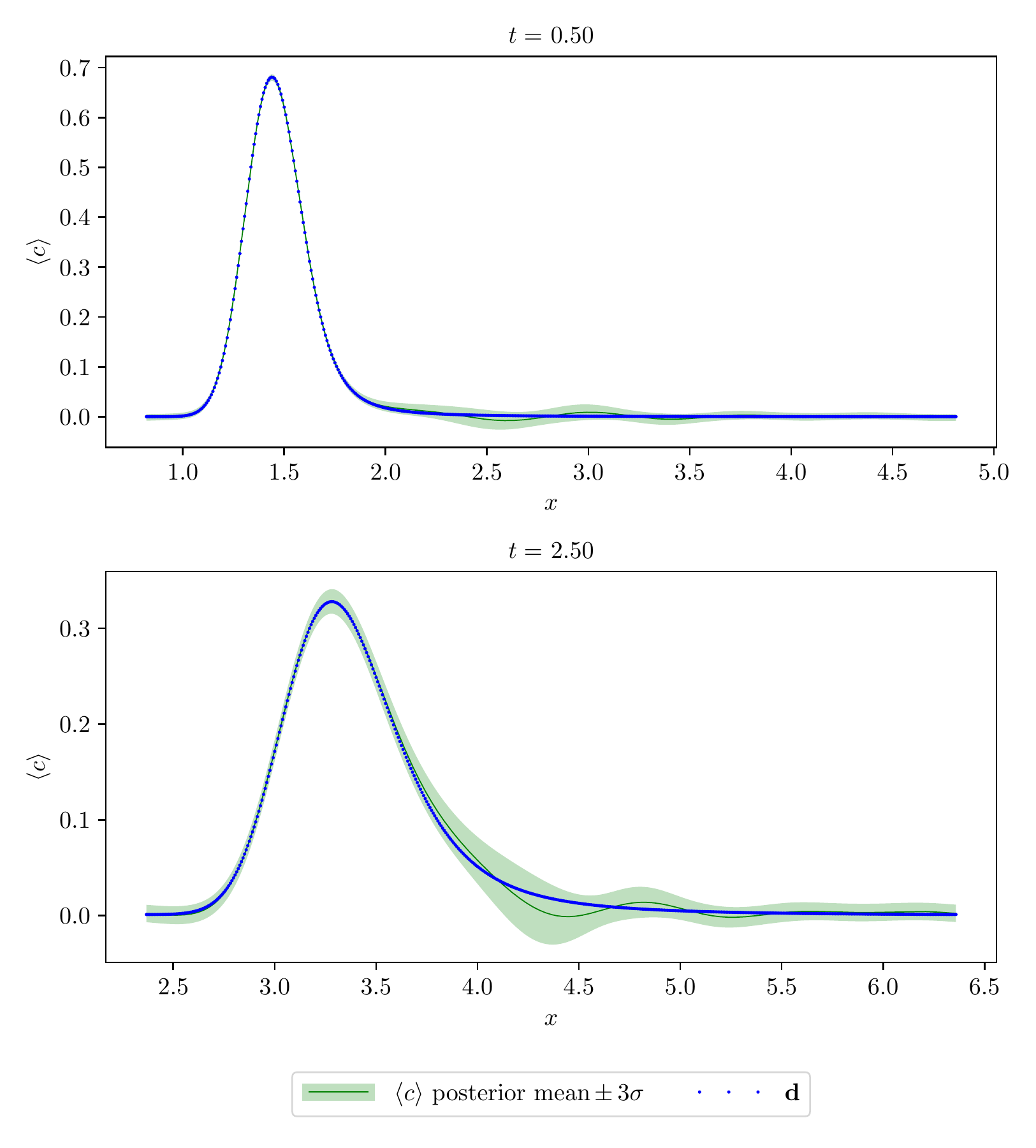}
  \caption{
    Statistics of $\mean{c}$ evolved using posterior samples from inference
    using 32 time-series observations of the solution of (\ref{eq:FRADE}) taken at
    $x=2.0$ and uniformly spaced over $[0,4]$ in time.}
  \label{fig:oscillating_push_forward}
\end{figure}

Each time series of $\mean{c}$ observations was sensitive to only the first 5 eigenvalues of $\mathcal{D}$. 
The time-series data may be sensitive to fewer eigenvalues than the spatial-series data because higher wavenumber modes are more rapidly damped, so that only data from the early times in the time series are sensitive to them.
Once again, increased frequency of observation increased information gain for the informed eigenvalues, as shown in \Cref{fig:kl_div_x_4.0}.
The information gain was less sensitive to the location at which the time-series data was collected than it was to the time at which spatial-series data was collected, as shown in \Cref{fig:time_kl_div_nobs_512}.
This may be due to the observation locations not being far enough apart to significantly change the information available from the time series.
While fewer eigenvalues were informed by time-series data than by spatial-series data, the information gain in those that were informed is similar. 
Both for time-series and spatial-series data, the posterior distributions for cases with abundant ($N_{obs}\!=\!512$) data contained the true value of the eigenvalues in their high-probability regions (see, e.g.~\Cref{fig:case1_n512_posteriors}).
  Note that the support of the posterior is so concentrated relative to the prior distribution that the prior appears flat at the scale of the posterior.
  Furthermore, statistics of $\mean{c}$ evaluated using posterior samples of $\Dcal$ and evolved outside the regime of the inference data were consistent with the true evolution, as shown in \Cref{fig:extrapolated_spatial_stats}.
For the sparsest data ($N_{obs}\!=\!32$), the posterior marginal distributions also largely contained the true value of the eigenvalues in their high-probability regions (see, e.g.~\Cref{fig:case1_n32_posteriors}).
  However, statistics of $\mean{c}$ evaluated using posterior samples of $\Dcal$ and evolved outside the regime of the inference data can yield 
nonphysical oscillations in the tails of $\mean{c}$ and negative concentrations, as shown in \Cref{fig:oscillating_push_forward}.

The likelihood can only penalize oscillations that induce large misfits with the data.
Sparse observations allow for oscillations to occur between the data points.
Time-series data is especially ill equipped to penalize oscillations, since they can occur anywhere in the spatial domain, as long as they are not evident when the solution crosses the observation point.
This can be seen, for instance, in \Cref{fig:oscillating_push_forward}, where oscillations in the tails are present downstream of the observation point at $x=2.0$. 
This result highlights the need to impose as many constraints as possible in data-sparse situations, especially when the available data cannot penalize a particular nonphysical behavior.
  The MLE FRADE solutions from the first-pass optimization used to seed the MCMC chains were observed to be positive in all data scenarios,
  so the oscillatory behavior and negative concentrations are induced when the set of sensitive eigenvalues are calibrated separate from the rest of the spectrum. 
  The positivity of the PDE solution depends on the spectrum as a whole, so it is postulated that a correlation relationship between the eigenvalues could be imposed which would guarantee positivity; however, such a relationship was not determined as part of this work. 
  Correlation matrices computed using posterior samples for the inference using $512$ spatial-series observations at time $0.5$ and for the inference using $32$ time-series observations at $x=2.0$ are presented for the interested reader in \Cref{fig:case1_n512_corr_mat} and \Cref{fig:case1_n32_corr_mat}, respectively.

\FloatBarrier

\section{Case 2: Data from the direct numerical computation of $\meanc$}\label{sec:case_2}
The study in \Cref{sec:case_1} indicates that it is possible to infer the eigenvalues to which $\mean{c}$ is sensitive, at least when the true operator lies in the probability space of the operator being inferred.
Based on this positive result, the same procedure is used in this section to infer the uncertain eigenvalues using data generated from the detailed evolution equations for $c$, \eqref{eq:detailedConsMass}-\eqref{eq:darcy}.
The full specification and implementation of this high-fidelity model is detailed in Appendices A, B, and C of \cite{portone_thesis}.
  Observations of $\mean{c}$ are produced by averaging over an ensemble of evolutions of $c$, generated by solving the high-fidelity model defined in \eqref{eq:detailedConsMass}-\eqref{eq:darcy} for an ensemble of independent, identically-distributed samples of the permeability field $\kappa$.
The goal of this study is to examine the spectrum of the uncertain linear differential operator inferred using data exhibiting anomalous diffusion, and compare its spectrum to that of the operators typically used as models of the phenomenon.
Whether the inferred spectrum is valid for times not included in the inference is also considered.

\subsection{Data and measurement error}
The data used for the inference was collected as an ensemble average of depthwise-averaged solutions of the high-fidelity model defined in \eqref{eq:detailedConsMass}-\eqref{eq:darcy} with initial condition
\begin{eqnarray*}
\begin{aligned}
  c_0(x,y) = \e{ -\frac{(1-x)^2}{2\beta^2} } \qquad\mbox{with $\beta=0.1$}
\end{aligned}
\end{eqnarray*}
on the domain $[0,4]\times[0,1]$.
The ensemble average was computed over the space of permeability fields distributed according to
\begin{equation*}
  \begin{aligned}
    \ln \kappa &\sim \Ncal( 0, C(\mathbf{x},\mathbf{x}') ), \\
    C(x,x') &\equiv \sigma^2 \e{ -\frac{(x - x')^2}{2\ell_x^2} - \frac{( y - y' )^2}{2 \ell_y^2} }.
  \end{aligned}
\end{equation*}
An ensemble of 576 evolutions of $c$ was generated by sampling $\kappa$ and solving \eqref{eq:detailedConsMass}-\eqref{eq:darcy}. 
These 2D evolutions were averaged in $y$, and the sample mean of these depth-averaged solutions $\mean{c}_y$ was computed and used as data for inference, denoted here $\mean{c}_N$, $N=576$.
Observations were taken at all 512 points on the grid in the streamwise direction at time $t=0.4$, or $1/10$ of a flowthrough time ($\sfrac{L_x}{\mean{u}}=4$).

\subsection{Likelihood}
The sampling error in the sample mean $\mean{c}_N$ was assumed to be consistent with the multidimensional Central Limit Theorem:
\begin{equation*}
  \begin{aligned}
    \dvec =  \mean{\cvec}_N + \eps, \quad\quad \eps \sim \Ncal\left( 0, \frac{\Sigma}{N} \right),
  \end{aligned}
\end{equation*}
where $\Sigma$ is the covariance matrix of the distribution of $\mean{c}$, which was estimated using the sample covariance matrix $S_N$.
In the tails of the ensemble-averaged concentration field, far from the mode of the pulse, the sample variance approached zero.
To avoid numerical issues in inverting the covariance matrix for the likelihood computation a minimum variance of $10^{-6}$ was imposed, yielding the likelihood
\begin{equation*}
  \begin{aligned}
    p( \mathbf{d} | \Thetavec ) &= \e{ -\frac{1}{2} \norm{ \mean{\mathbf{c}}( \Thetavec ) - \dvec }_{S^{-1/2}} }, \\
    S_{ij} &= \fndef{
      \max\left( \sfrac{(S_N)_{ij}}{N}, 10^{-6} \right), & i=j, \\
      \sfrac{(S_N)_{ij}}{N}, & i\not=j.
    } 
  \end{aligned}
\end{equation*}

\subsection{Results}\label{sec:case2_results}
\begin{figure}[h]
\centering
\includegraphics[scale=.9]{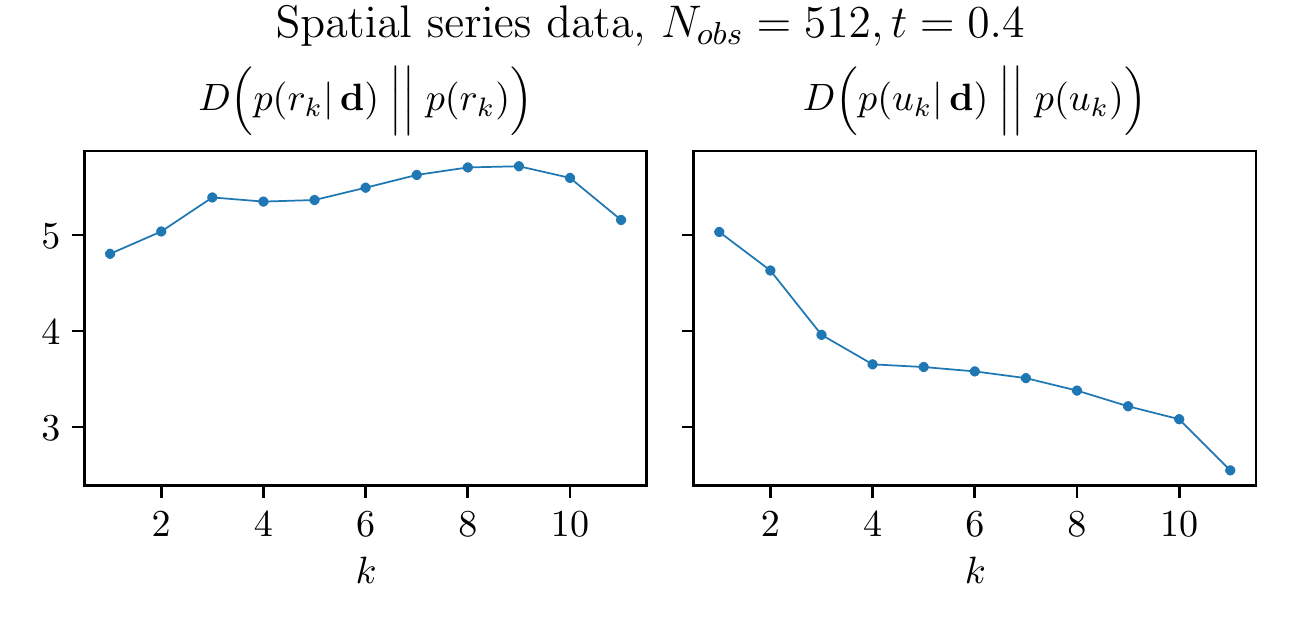}
\caption{KL divergence of posteriors relative to priors for Bayesian
  inference from spatial-series data generated from the high-fidelity model.}
\label{fig:case2_kl_divs}
\end{figure}

\begin{figure}[h]
\centering
\includegraphics[scale=.9]{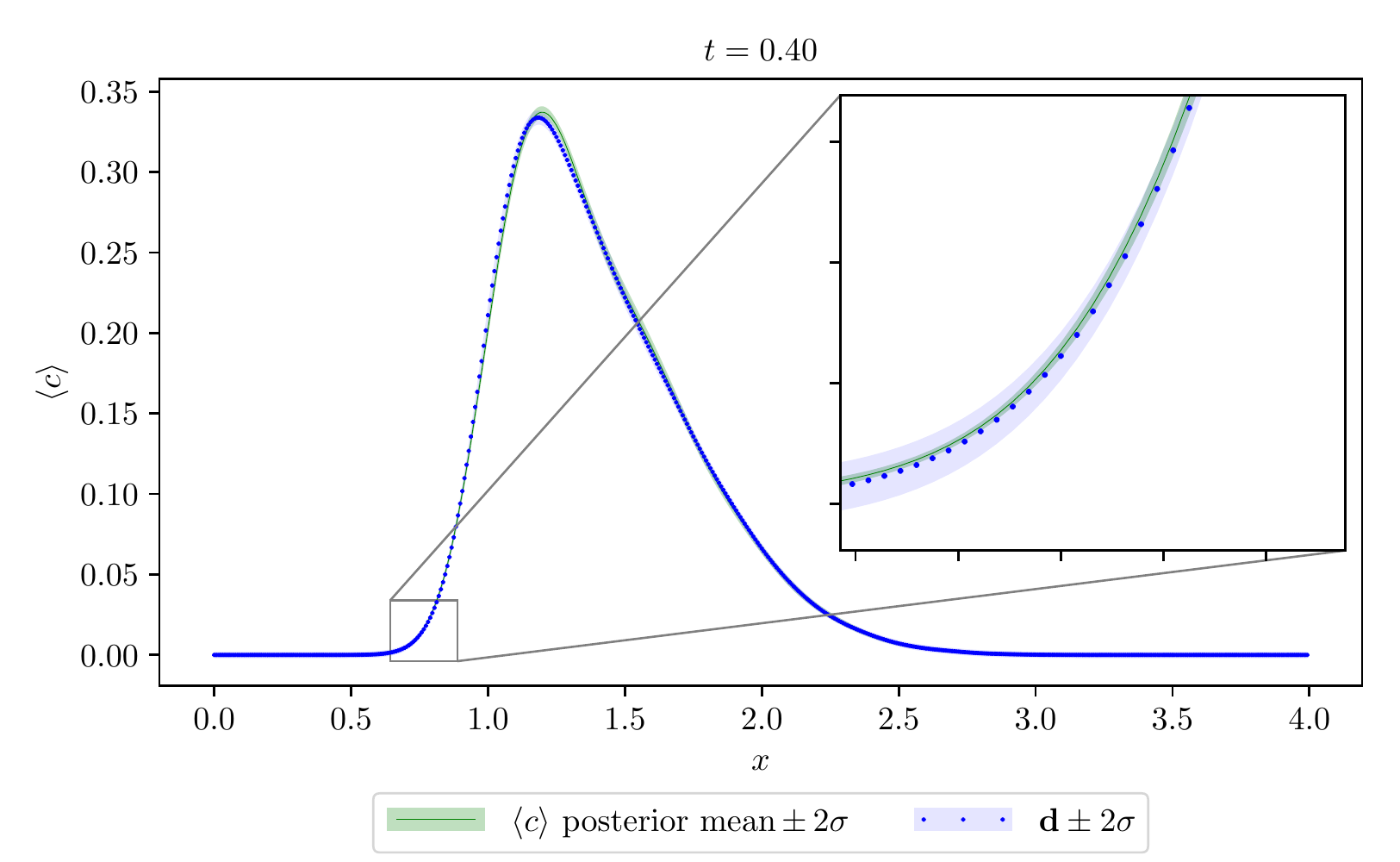}
\caption{
  Evolutions of $\mean{c}$ evaluated using posterior samples inferred from
  spatial-series data with 512 observations of the high fidelity model at
  $t=0.4$, compared to the inference data.
}
\label{fig:case2_function_stats}
\end{figure}

\begin{figure}[h]
\centering
\includegraphics[scale=.9]{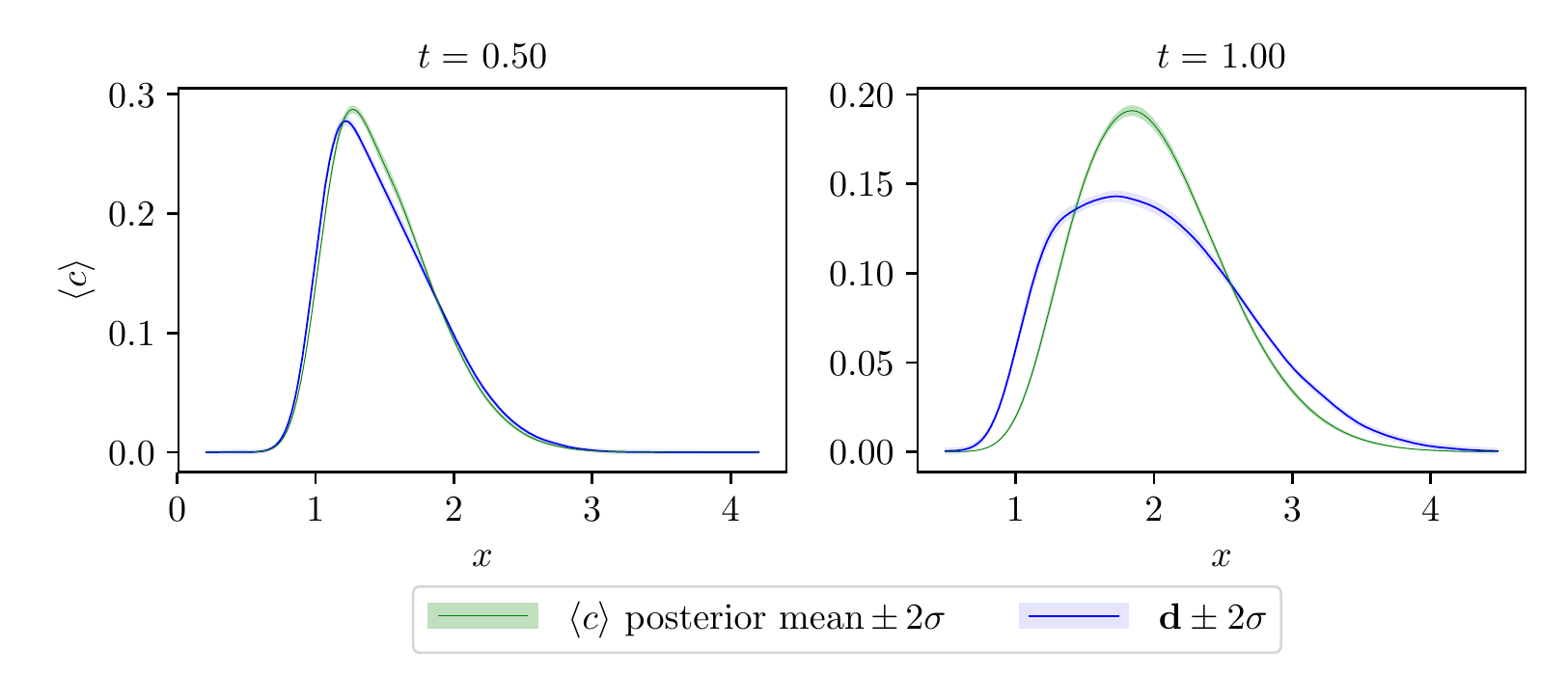}
\caption{
  Evolutions of $\mean{c}$ at $t=0.5$ and $1.0$ evaluated using posterior samples inferred from
  a spatial-series of 512 observations of the high fidelity model at
  $t=0.4$, compared to the high-fidelity model.
}
\label{fig:case2_extrapolated_stats}
\end{figure}

The eigenvalues of $\Dcal$ were inferred using abundant spatial data to replicate the most successful conditions for inference from \Cref{sec:case_1}.
  This abundance of spatial observations penalizes nonphysical oscillations in $\mean{c}$ through the likelihood, since the operator is not constrained against this behavior directly.
  As with \Cref{sec:case_1}, it is expected that if sparse time-series data were used, the same oscillatory behavior could arise. 
  However, the data in this case is smoother than in \Cref{sec:case_1}, which was corrupted with uncorrelated noise; with sparse data the likelihood in \Cref{sec:case_1} would be minimized for evolutions of $\mean{c}$ that had small fluctuations that fit the noise in the data, while such oscillations in the data do not occur in this case. 
  Because of this it is possible that in this case the oscillations in $\mean{c}$ would be much less significant relative to \Cref{sec:case_1}, though this possibility was not explored here.
The solution $\mean{c}$ from the generalized ADE was sensitive to the first 11 eigenvalues, using the same sensitivity analysis procedure as \Cref{sec:case_1}.
The KL divergences for the informed eigenvalues are shown in \Cref{fig:case2_kl_divs}.
The KL divergences are lower than in \Cref{sec:case_1}, presumably because the correlation in the data makes it less informative.
  The correlation matrix computed using posterior samples is presented for the interested reader in \Cref{fig:case2_corr_mat}.
  The resulting evolutions of $\mean{c}$ evaluated using posterior samples indicate good agreement with the data used in the inference, as shown in \Cref{fig:case2_function_stats}.
However, the solution for $\meanc$ at later times, as in \Cref{fig:case2_extrapolated_stats}, makes it clear that the inferred eigenvalues for $\Dcal$ do not capture the evolution of $\mean{c}$.
To reproduce the evolution of $\mean{c}$ the eigenvalues of $\Dcal$ must be time dependent.
Since $\Dcal \equiv \nu_p \sfrac{\partial^2}{\partial x^2}+ \Lcal $, the only possible source of this time dependence is $\Lcal$.
As discussed in \Cref{sec:operator_formulation}, the eigenvalues of the deterministic operator $\Lcal$ that would reproduce the effects of dispersion on the mean are time dependent, so this does not come as a surprise.

\begin{figure}[h]
  \centering
  \includegraphics[scale=0.9]{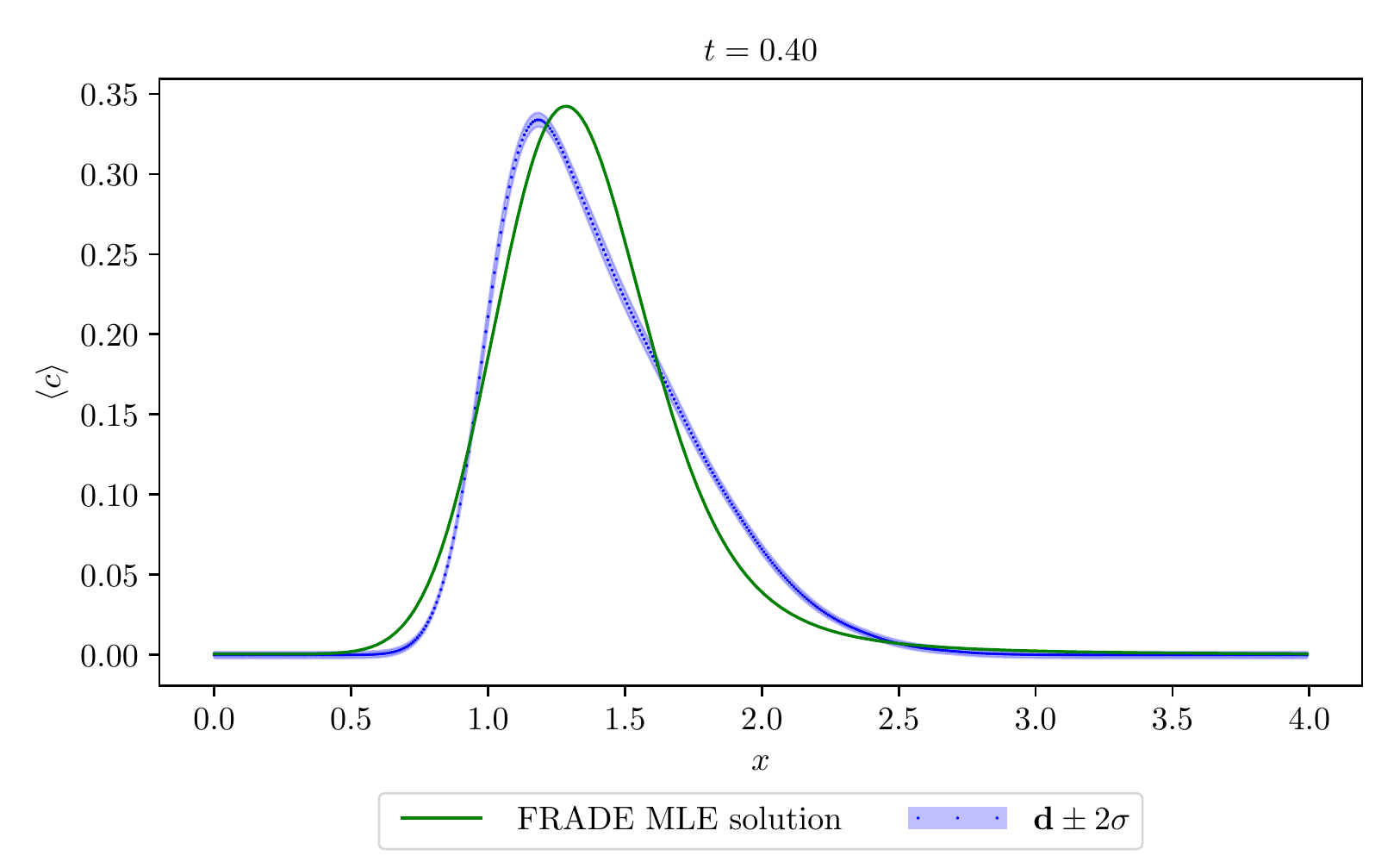}
  \caption{FRADE solution $\mean{c}$, using the maximum likelihood
    estimator values of $\nu$ and $\alpha$ for the fractional
    derivative, compared to the calibration data from the high-fidelity model. }
  \label{fig:frade_opt_vs_data}
\end{figure}

\begin{figure}[h]
  \centering
  \includegraphics[scale=0.9]{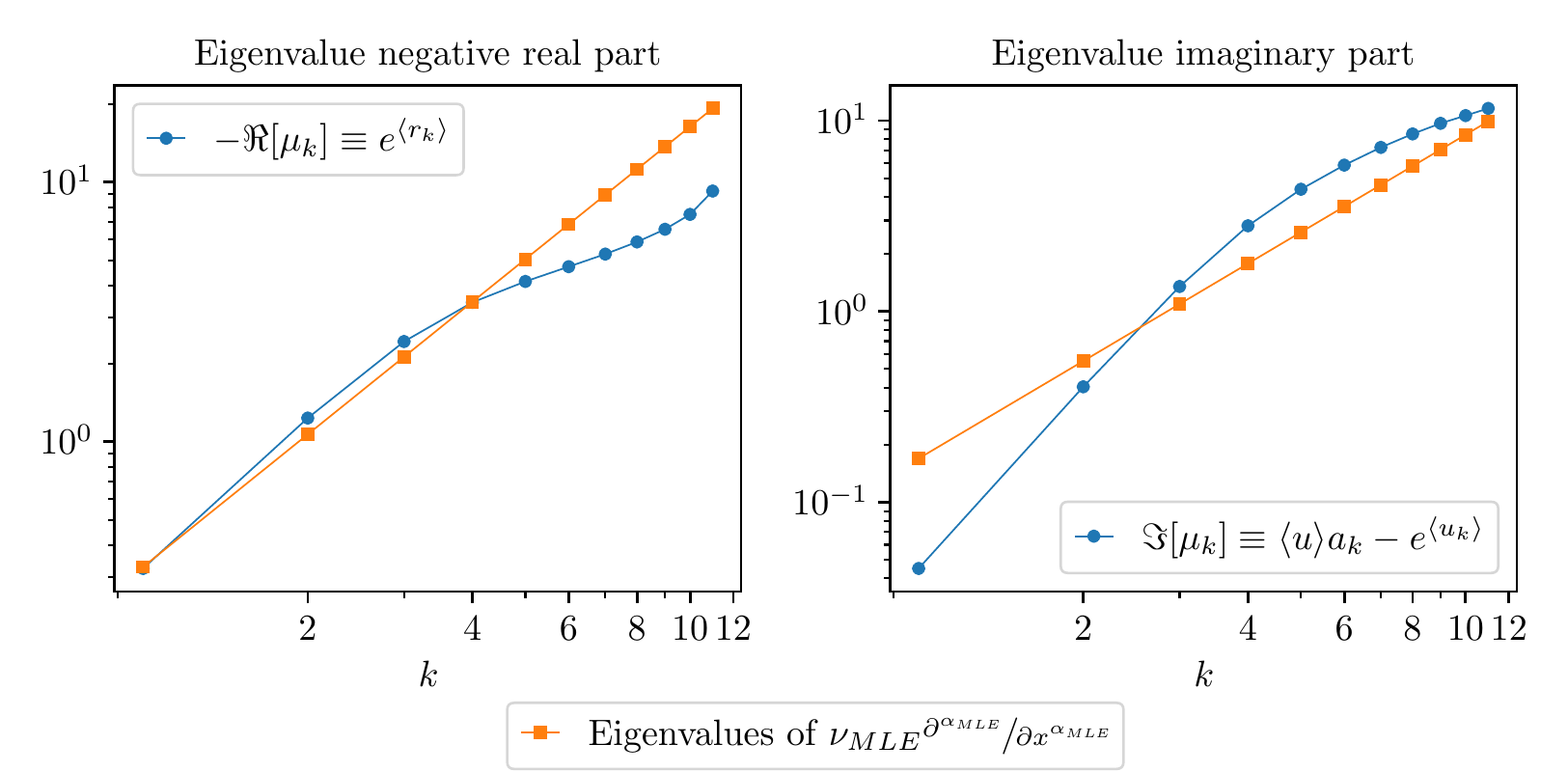}
  \caption{The real and imaginary parts of $\mu_k$,
    computed using the posterior mean values of $r_k$ and $u_k$,
    compared to the eigenvalues of the fractional derivative with maximum likelihood
    estimator values of $\nu$ and $\alpha$.}
  \label{fig:posterior_mean_vs_frade}
\end{figure}

Though it cannot successfully extrapolate in time, 
the operator was general enough that its posterior was consistent with the data used to infer its eigenvalues.
In comparison, the fractional derivative that maximized the likelihood (the FRADE MLE) was not consistent with the data, as shown in \Cref{fig:frade_opt_vs_data}.
As shown in \Cref{fig:posterior_mean_vs_frade}, the mean of the inferred spectrum of $\Dcal$ exhibits a more complex dependence on wavenumber than would be captured with a fractional derivative. 
While the magnitude of the fractional derivative eigenvalues grows as a fixed power of $k$, this is not true for the inferred eigenvalues.
Additionally, the rate of growth as a function of $k$ is different between the real and imaginary parts of the eigenvalues.
Finally, note that the imaginary parts of the inferred eigenvalues are the same order of magnitude as the real part, in contrast to a gradient-diffusion model of dispersion which predicts real eigenvalues.
These findings indicate that more complex wavenumber dependence, as well as time dependence, are important to the development of an adequate closure model for anomalous diffusion.

\FloatBarrier

\section{Conclusions}\label{sec:conclusions}
In this paper a Bayesian inverse problem was posed to infer the spectrum of an infinite-dimensional differential operator appearing as a closure term in a model for mean contaminant transport through a heterogeneous porous medium.
Observations of the state variable on which the operator acts were used as inference data.
The operator was parametrized using its eigendecomposition, and physics-based constraints were mathematically imposed on its eigenfunctions and eigenvalues.
In this case the operator's eigenfunctions were known to be the Fourier modes. 
Remaining uncertainties in its eigenvalues were represented as probability distributions, which were updated using Bayes's theorem.

The parameterization of the operator using its eigendecomposition provided a useful insight into its action on the state variable.
Most of the relevant physical constraints in the problem translated into straightforward constraints on the eigendecomposition.
However, a simple, constructive method to enforce positivity preservation on the operator was not available. 
This is presumably because of the nonpositivity of the Fourier mode eigenfunctions, rather than because of the infinite-dimensional operator formulation itself.
As shown in \Cref{sec:case_1}, with enough spatial data to penalize nonphysical behavior, the lack of this constraint can be overcome to infer an operator that produces physical evolutions. 
However, for data that does not penalize nonphysical behavior, such as sparse time-series data, the constraint would be necessary.
This demonstrates the importance of enforcing as much prior information as possible in scenarios with limited data.

In the case where the uncertain operator could exactly represent the underlying dynamics of the problem, as in \Cref{sec:case_1}, the inferred eigenvalues converged to their true values with increasing frequency of observations in time or space.
The inherent dimension of the Bayesian inverse problem is limited by the spectral content of the state, as was illustrated through the global sensitivity analysis performed in the different data scenarios.
This dimension is independent of the discretization of the problem.
Because of this, although the operator is defined to be infinite dimensional, the effective dimensionality of the problem in all cases was relatively small, not exceeding 10 eigenvalues in any of the data scenarios studied.

In \Cref{sec:case_2}, the operator's eigenvalues were inferred using data generated from a high-fidelity model that exhibits anomalous diffusion. 
The operator's eigenvalues were inferred using observations of the state variable at a single time and at every point in the computational domain. 
Though the model evaluations using samples from the posterior of the inferred operator were consistent with the calibration data, they were not consistent with observations of the state at later times.
This is not surprising since, as discussed in \Cref{sec:operator_formulation}, the operator's eigenvalues would need to be time dependent to fully capture the time-dependent relationship between $\mean{c}$ and $\mean{u'c'}$.
The operator formulation posed here is more general than the fractional-derivative and gradient-diffusion models which are common closures for dispersion, since it does not require the eigenvalues grow according to fixed power of $k$. 
However, even this more general operator could not successfully extrapolate in time.
This suggests that a successful closure representation of anomalous diffusion must account for the time dependence of the process, perhaps through use of a richer state description e.g.~modeling the evolution of the variance of $c$.

This work is an initial step in assessing the feasibility of inferring an uncertain operator appearing in a PDE-based physical model using limited data.
It does not address the well-posedness of the infinite-dimensional inference problem, however the generalization of existing theory for infinite-dimensional Bayesian inverse problems to operators is an interesting future research direction.
The inverse problem was cast in terms of its eigendecomposition, which enabled physical constraints to be encoded deterministically and in a straightforward manner.
  Additionally, qualitative physical information was encoded through the prior distribution used in the inference.
The eigendecomposition formulation of the inverse problem exposed the inherent dimensionality of the problem, based on the number of eigenvalues which were informed by the data, which is independent of discretization.
Given the generality of the operator's form, the importance of enforcing any known physical constraints in cases of sparse data is essential to achieving physically meaningful results.
Nevertheless, the success in inferring the operator's spectrum in \Cref{sec:case_1} suggests this approach is promising.

  There are several potential extensions to this formulation.
  First, a straightforward extension to a three-dimensional high-fidelity model with two homogeneous directions would be possible using the presented eigendecomposition parameterization.
  The eigendecomposition approach is not limited to stochastically-upscaled models; it can be applied to any model in which an invariance to a continuous tranformation exists and can be exploited to determine the eigendecomposition of the operator.
  In this case translation invariance was exploited to identify the Fourier modes as the eigenfunctions of the operator, but systems with rotational invariance would admit an eigendecomposition in terms of spherical harmonics.
  More generally, the approach of augmenting limited data with qualitative physical information through the prior distribution of a tractably-parametrized uncertain operator is applicable in the case of nonlinear problems as well, including nonlinear uncertain operators.
  For weakly nonlinear problems it may be possible to employ the given formulation on the linearized system along with a low-dimensional parametrization of a nonlinear correction.
  For strongly nonlinear problems, methods to transform nonlinear equations into more tractable forms using  variable transformations and introductions of auxiliary variables, as in \cite{gu2011} and \cite{kramer2019}, could yield tractable nonlinear operators that are amenable to imposing prior physical constraints.
  For example, prior constraints were placed on a low-dimensional nonlinear operator in \cite{morrison}, although the operator was finite-dimensional rather than infinite-dimensional as was considered here.

\section{Acknowledgements}
The support of this work by the U.S.~Department of Energy, carried out at the University of Texas at Austin under contracts DE-SC0009286 and DE-SC0019303, is gratefully acknowledged. 
This paper describes objective technical results and analysis. Any subjective views or opinions that might be expressed in the paper do not necessarily represent the views of the U.S. Department of Energy or the United States Government.
The authors acknowledge the Texas Advanced Computing Center (TACC) at The University of Texas at Austin for providing HPC resources that have contributed to the research results reported within this paper. URL: http://www.tacc.utexas.edu.

\appendix
\section{Posterior correlation matrices}
\begin{figure}[h]
  \centering
  \begin{subfigure}[b]{0.49\textwidth}
    \centering
    \includegraphics[scale=0.75, trim=2em 0 0 0 ]{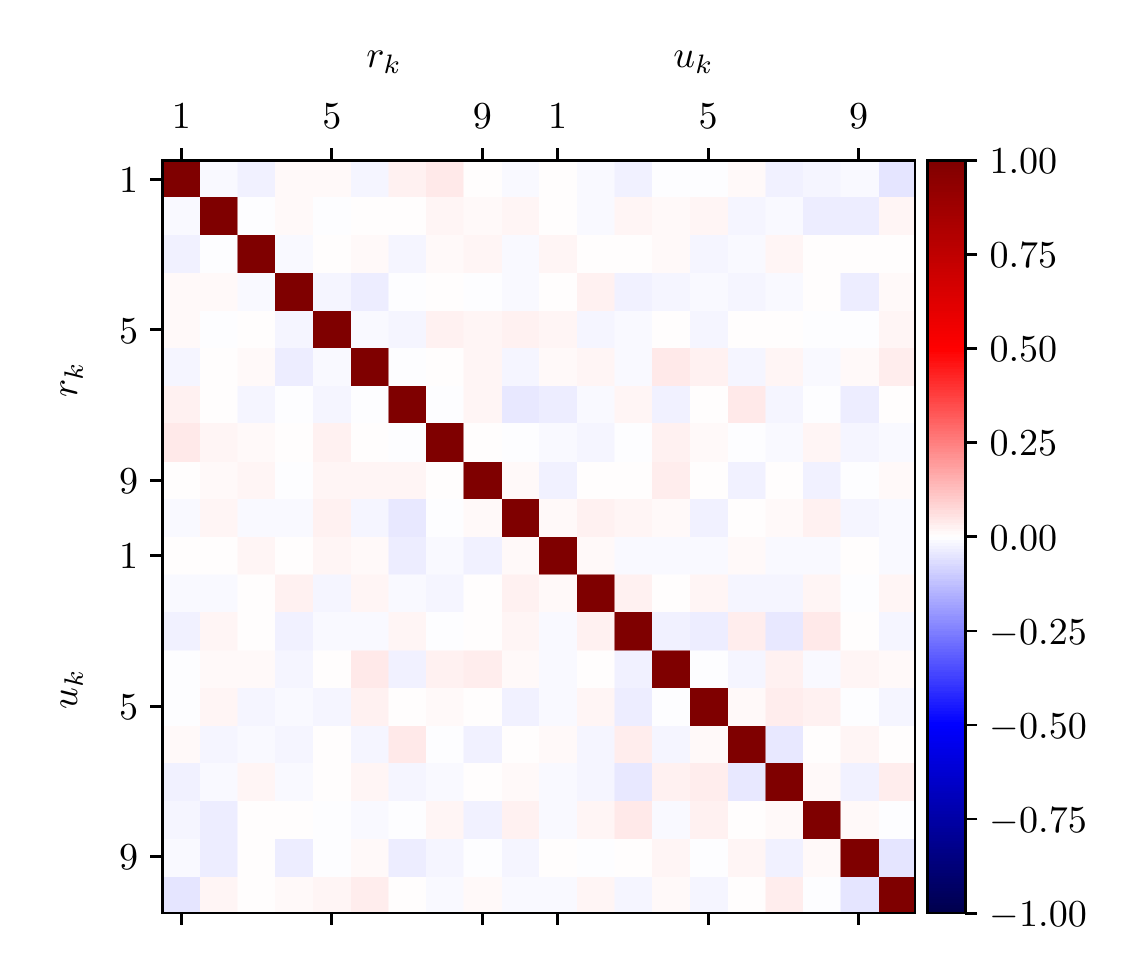}
    \caption{Correlation matrix of posterior samples from inference with 512 spatial observations of the solution to \eqref{eq:FRADE} equally spaced over the spatial domain, taken at time $t=0.5$. }
    \label{fig:case1_n512_corr_mat}
  \end{subfigure}
  \hfill
  \begin{subfigure}[b]{0.49\textwidth}
    \centering
    \includegraphics[scale=0.75, trim=2em 0 0 0 ]{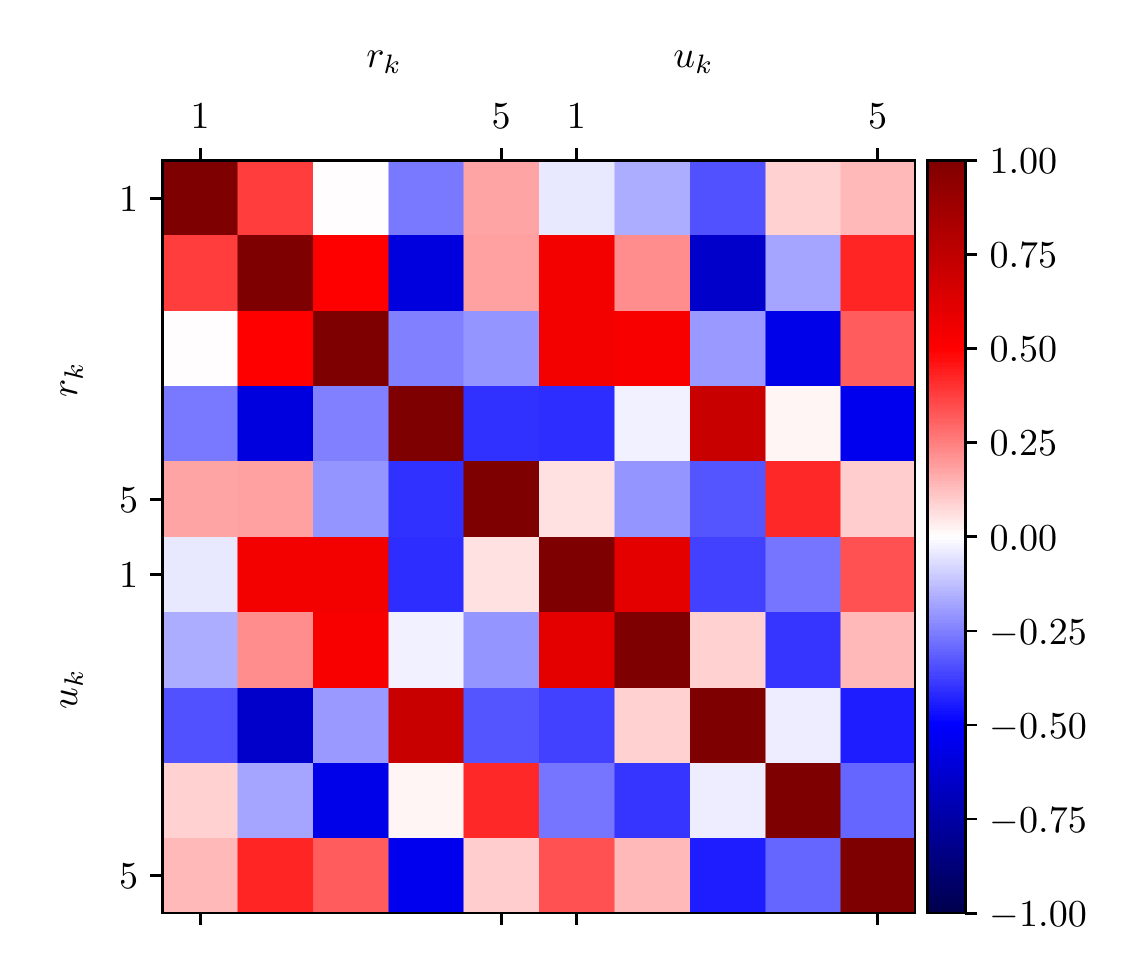}
    \caption{Correlation matrix of posterior samples from 
    inference using 32 time-series observations of the solution of \eqref{eq:FRADE}
    taken at $x=2.0$ and uniformly spaced over $[0,4]$ in time.}
    \label{fig:case1_n32_corr_mat}
  \end{subfigure}
  \begin{subfigure}[b]{0.9\textwidth}
    \centering
    \includegraphics[scale=0.75, trim=2em 0 0 0 ]{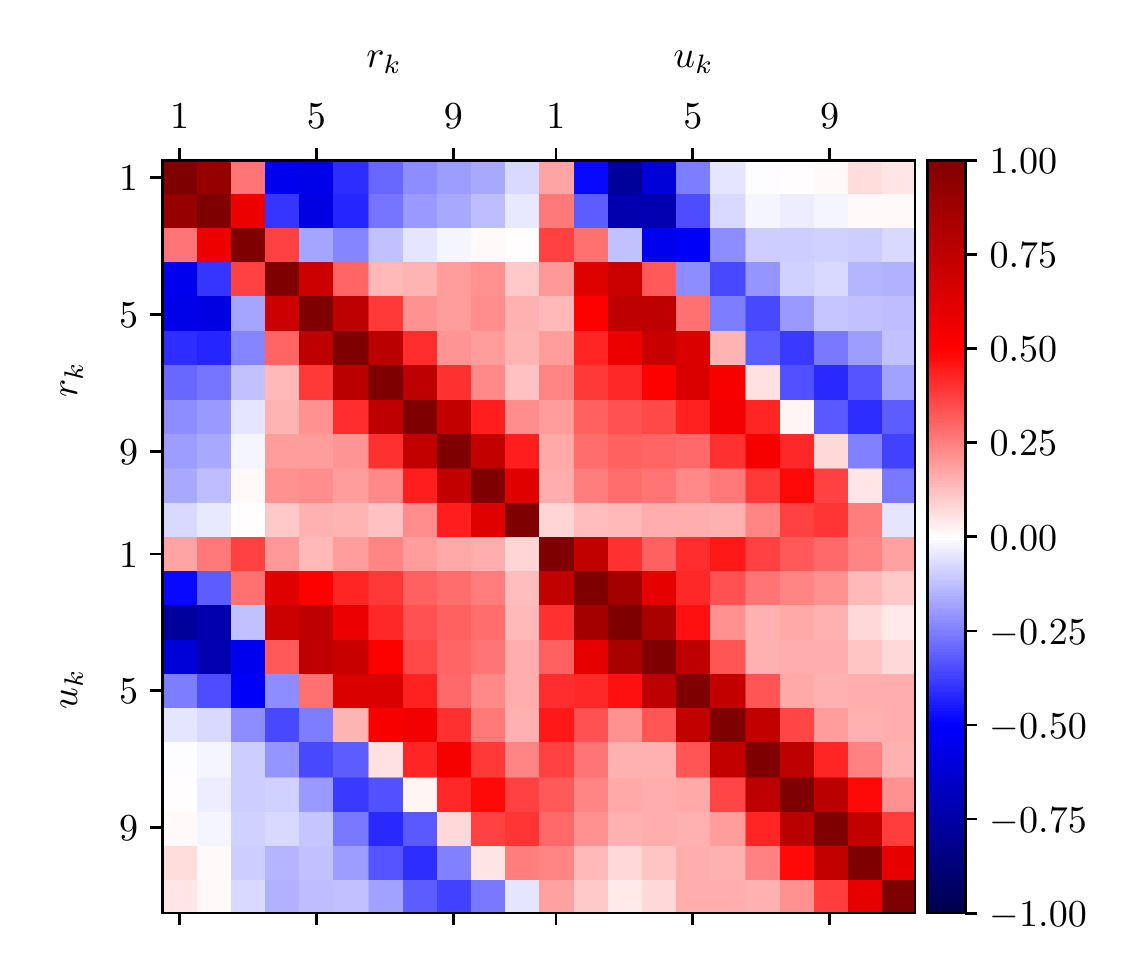}
    \caption{Correlation matrix of posterior samples from 
      the inference defined in \Cref{sec:case_2}. 
  }
    \label{fig:case2_corr_mat}
  \end{subfigure}
  \caption{Correlation matrices computed using posterior samples from the three inference scenarios discussed in detail in \Cref{sec:case_1,sec:case_2}. The upper-left quadrant of the correlation matrix contains correlations between the real parts of the eigenvalues, the bottom-left quadrant contains the correlations between the imaginary parts, and the top-right and bottom-left quadrants contain the correlations between the real and imaginary parts.
  }
  \label{fig:corr_mats}
\end{figure}

Posterior correlation matrices for the three cases discussed in detail in \Cref{sec:case_1,sec:case_2} are presented in \Cref{fig:corr_mats}.
Interestingly, the structure of the correlation matrices is quite different between the three cases.
The correlation matrices for all spatial-series cases resemble \Cref{fig:case1_n512_corr_mat}, regardless of observation frequency or time of observation.
Similarly, the correlation matrices for all time-series cases resemble \Cref{fig:case1_n32_corr_mat} regardless of observation frequency or location of observation.
The posterior correlation matrix for the inference performed in \Cref{sec:case_2} using ensemble-averaged observations of the high-fidelity model \eqref{eq:detailedConsMass}-\eqref{eq:darcy} also exhibits strong positive and negative correlations between the eigenvalues, even though spatial-series observations were used in this case.
The causes and interpretation of the differences in the posterior correlations across these cases are left for future study.
\FloatBarrier


\bibliographystyle{siamplain}
\bibliography{references}

\end{document}